\documentclass[11pt,twocolumn,twoside]{osajnl}

\usepackage{lineno,hyperref}

\usepackage{geometry}
\usepackage{graphicx}
\usepackage{amsmath, amsthm, amssymb, amsfonts}
\usepackage{epstopdf}
\usepackage{subfigure}
\usepackage{algorithm}
\usepackage{algpseudocode}
\usepackage{amsmath,amssymb,graphicx}
\usepackage{tabu}
\usepackage{caption}
\usepackage{xcolor}
\usepackage{textcomp}

\long\def\comment#1{}

\newcommand{\Hop}{\mathop{\mathcal H}}

\newcommand{\argmin}[1]{\underset{#1}{\operatorname{argmin}}\;}

\long\def\ie{\emph{i.e.}} 
 
\long\def\eg{\emph{e.g.}}

\journal{ao} % Choose journal (ao, aop, josaa, josab, ol, pr)

\setboolean{shortarticle}{false}
\title{Deep neural network for fringe pattern filtering and normalisation}

\author[1]{Alan Reyes-Figueroa}
\author[1,*]{Mariano Rivera}

\affil[1]{Centro de Investigaci\'on en Matem\'aticas AC, 36023, Guanajuato, Gto., M\'exico}

\affil[*]{Corresponding author: mrivera@cimat.mx}

\ociscodes{(100.0100) Image processing;  (100.4996)  Pattern recognition, neural networks;   (100.2650) Fringe analysis; (100.3008)   Image recognition, algorithms and filters; (120.6165)   Speckle interferometry, metrology}

\doi{\url{http://doi.org/XX.XXXX/XX.XX.XXXXXX}}

\begin{abstract}
We propose a new framework for processing Fringe Patterns (FP). Our novel approach builds upon the hypothesis that the denoising and normalisation of FPs can be learned by a deep neural network if enough pairs of corrupted and ideal FPs are provided. The main contributions of this paper are the following: (1) We propose the use of the U--net neural network architecture for FP normalisation tasks; (2) we propose a modification for the distribution of weights in the U--net, called here the V--net model, which is more convenient for reconstruction tasks, and we conduct extensive experimental evidence in which the V--net produces high--quality results for FP filtering and normalisation. (3) We also propose two modifications of the V--net scheme, namely, a residual version called ResV--net and a fast operating version of the V--net, to evaluate the potential improvements when modify our proposal. We evaluate the performance of our methods in various scenarios: FPs corrupted with different degrees of noise, and corrupted with different noise distributions. We compare our methodology versus other state-of-the-art methods. The experimental results (on both synthetic and real data) demonstrate the capabilities and potential of this new paradigm for processing interferograms.
\end{abstract}

\setboolean{displaycopyright}{true}

\begin{document}

\maketitle

%------------------------------------------------------------------------------------------------------------------------
\section*{1.  Introduction}
%------------------------------------------------------------------------------------------------------------------------

Fringe Pattern (FP) denoising--normalisation consists of removing background illumination variations, normalising amplitude and filtering noise, which means transforming an FP corresponding to the mathematical model
\begin{equation}
 	\label{eq:fp}
	x(p) = a(p) + b(p) \cos\left( \phi(p) \right) + \eta(p)
\end{equation}
into the normalised FP modelled by
\begin{equation}
 	\label{eq:fpnorm}
	\hat x(p) = 1 + \cos\left( \phi(p) \right).
\end{equation} 
Here,  $x$ is the signal or irradiance, $a$ the background illumination, $b$ the fringe amplitude or modulation, $\phi$ the phase map, $\eta$ is an additive or correlated noise, and $p$ denotes the pixel position, to indicate the spatial dependence of those quantities. Such a normalisation can be represented by the transformation
\begin{equation}
 	\label{eq:normalise}
	\hat x = \Hop \{ x \}.
\end{equation}

FP normalisation is a critical step of the fringe analysis processing. There is a consensus that FP analysis can be seen as a pipeline that involves the following steps: denoising, normalisation, phase extraction and phase unwrapping, even if some of these steps are merged by some methods. In all these steps, background and modulation variations are considered as a source of error in the phase estimation \cite{quiroga:03}. More precisely, in the most simple version of phase estimation \cite{quiroga:01}, one introduces a set of $n \geq 2$ known phase steps $\delta_k$, $k = 1,2,\ldots, n$ in the modulating phase and a set of signals
\begin{equation}
 	\label{eq:steps}
	x_k(p) = a_k(p) + b_k(p) \cos\left( \phi(p) + \delta_k \right) + \eta_k(p),
\end{equation}
$1 \leq k \leq n$, are acquired.

The wrapped phase is estimated by
\begin{equation}
 	\label{eq:tangent}
		\hat \phi(p) = \arctan_2 \big(x_s(p), x_c(p) \big),
\end{equation}
where $x_s$ and $x_c$ denote the signals associated to the sine and cosine, respectively, of the phase multiplied by the modulation amplitude, that is
\begin{equation}
 	\label{eq:components}
	x_s(p) = b(p) \sin \phi(p); \ \ \  x_c(p) = b(p) \cos \phi(p).
\end{equation}
Both terms are always computed as a combination of the phase shifted images $x_k$, $1 \leq k \leq n$.

It is well known that the quality in the approximation represented by the components (\ref{eq:components}) determines the error of the phase obtained by (\ref{eq:tangent}). Most phase estimation algorithms assume that the background $a$ and the modulation $b$ do not change with the phase step. When this requirement is not satisfied, \eg \ when there is a lack of illumination control, using directly equation (\ref{eq:tangent}) conduce to big errors in the phase estimation \cite{quiroga:01}. %demodulated and unwrapped phase estimation \cite{quiroga:01}.

It is well known that algorithms for phase demodulation from a single fringe pattern with closed fringes  \cite{marroquin:97, servin:98, servin:01, rivera:05} 
are particular sensitive when process noisy and denormalized fringe patterns: deviations in the normalisation $\hat x$ may produce failures in the phase estimation \cite{quiroga:01}.  Moreover, according to Ref. \cite{Flores:2020}, this problem also appears when analyzing pairs of fringe patterns with a phase shift.

Due to the reasons above, it is important to develop robust methods to filter the signal $x$ before the demodulation process.

Refs. \cite{gorthi:fp10, juarez:fprev15} present useful reviews of challenges related to FP analysis and a good list of proposed solutions.  For example, the denoising, the normalisation and the phase extraction can be accomplished using the two-dimensional Windowed Fourier Transform (WFT) \cite{kemao2007two}, the Wavelet Transform (WT), \cite{huang:waveletFP10, zhang:wft12} or a Gabor Filter Bank (GFB) based method \cite{rivera:twostep16}. As is noted in Refs. \cite{huang:waveletFP10, zhang:wft12}, WFT and FT have limitations in dealing with FPs when phase discontinuities are present or the field of view is not the full image. As we will show in this work, the same limitations apply for the GFB approach. These techniques estimate a transformation of the form \eqref{eq:normalise}, for a central pixel in an image neighbourhood (image patch or weighted image window).

% ------------------------------------------------------------------------------------------------------------------------------
\begin{figure}[h!]
 \centering \centering \includegraphics[width=\linewidth]{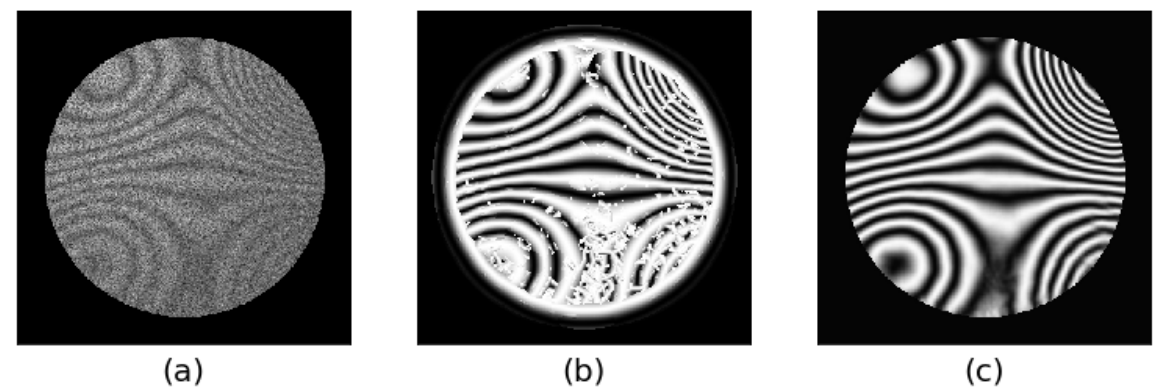}
 \caption{Normalisation of an FP with incomplete field of view: (a) data, (b) GFB (note the artefacts in regions with low--frequency fringes and near the border of the region of interest) and (c) proposal.}
 \label{fig:pupil}
 \end{figure}
% ------------------------------------------------------------------------------------------------------------------------------

WFT, WT and GFB methods rely upon the assumption that the neighbourhood of a pixel of interest (image patch) has an almost constant frequency and phase, \ie,  locally, the phase map is close to being a plane. The limitation of the mentioned methods occur at patches were the main assumption is violated; \ie, at phase discontinuities. Figure \ref{fig:pupil} shows a denoised-filtered FP computed with a GFB based method and with our proposal. Further information on alternative strategies for FP normalisation can be found in Ref. \cite{juarez:fprev15}. However, methods based on local spectral analysis (\eg, WFT, WT and GFB) have shown to be very robust general methods for dealing with high noise levels \cite{huang:waveletFP10, zhang:wft12, rivera:twostep16, dalmau:twostep16, rivera:transient18}. In this work, we propose to implement such a transformation \eqref{eq:normalise} with a Deep Neural Network (DNN).

Neural Networks (NNs) are known for being universal approximators \cite{kurkova92:approxNN}. 
Early attempts to use a NN for FP analysis are of limited application since they are based on simple multilayer schemes. In particular, in Ref. \cite{cuevas00:nnfringes} a multilayer NN is trained for computing the phase and its gradient at the central pixel of an image patch. Instead, our proposal computes the normalisation for the entire image patch. In addition, our work is based on a deep auto--encoder NN that allows us to deal with noise and large illumination changes. 

This paper investigates the V--net model performance in the task of normalise FPs and a comparison against other NN models is presented. Recently, we reported in Ref. \cite{Flores:2020} how the V--net performs in the task of normalise FPs for the phase step estimation. In such work, the V--net performance is favourably evaluated versus GFB and Hilbert-Huang Transform (HHT) methods.

Now, in this work, we present the details of the V--net. This paper is organized as follows: in Section 2 we present a brief review of the theoretical foundations of neural networks autoencoders, and the architecture details of our U--net variants: V--net and Residual V--net models. In Section 3 indicate the implementations details for the simulated data and the inference process for the normalized FPs using the V--net models. Section 4 is devoted to apply the V--net models to normalise synthetic FPs and evaluate the method in several scenarios. Conlusions are given in Section 5.

%------------------------------------------------------------------------------------------------------------------------
\section*{2.  Method}
\label{sec:method}
%------------------------------------------------------------------------------------------------------------------------

%\subsection*{2.1  Auto--encoders}
%\label{ssec:autoencoders}

\subsection*{2.1  The U--net model}
\label{ssec:unet}

Auto--encoders were originally devised to compress (codify) and decompress (decodify) data vectors \cite{Hinton504}. Auto--encoders have two main components: (1) The encoder $\mathcal{E}$ takes the data $x$ in its original ``spatial'' dimension and produces a compressed vector $y$. This encoding can be expressed by
\begin{align}
	\label{eq:encoder}
	y = \mathcal{E}(x) \overset{def}{=} \varphi_1 (W_1 x + b_1),
\end{align}
where $x$ is the original data, $W_1$ the weights matrix, $b_1$ a bias term, $y$ the encoded data and $\varphi_1$ is an activation function; \eg, ReLU, sigmoid or softmax \cite{nair10:relu}. (2) The decoder $\mathcal{D}$ takes the compressed data $y$ and computes a reconstruction $\hat x$ of the original data $x$, of the same dimension as $x$. This is expressed by
\begin{align}
	\label{eq:decoder}
	\hat x = \mathcal{D}(y) \overset{def}{=} \varphi_2 (W_2 y + b_2),
\end{align}
where $W_2$ is a weights matrix, $b_2$ a bias term and $\varphi_2$ is the activation function.

%%------------------------------------------------------------------------------------------------------------------------
%\subsection*{2.2  The U--net and V--net models}
%\label{ssec:method}
%%------------------------------------------------------------------------------------------------------------------------

In this work we propose to use a deep auto--encoder, called the U--net \cite{ronnenberg:Unet15}, and its here proposed variant V--net,  for performing FP denoising and normalisation. The U--net is a fully convolutional neural network (CNN), and was initially designed for image classification (segmentation). That is, for each input image, it produces an image of labels of the same size as the input \cite{fullyconvnet}, unlike standard classification networks whose output is a single value. The loss function for the U--net is of the form
 
\begin{align}
	\label{eq:autoencoder}
	\argmin{\theta} & \sum_i \| f(x_i) - (\mathcal{D} \circ \mathcal{E}) x_i \|_M,
\end{align}
where $f(x_i)$ is a segmentation of $x_i$, $\mathcal{E}$ and $\mathcal{D}$ represent the encoder and decoder stages, respectively; $\theta$ is the vector of the auto--encoder parameters, and $\| \cdot \|_M$ is a metric or divergence measure.

One can note important differences between the classical auto--encoders and the U--net: (1) U--net is a deep model; \ie, the number of layers in U--net is substantially larger than the number of layers in the classic auto--encoder. (2) U--net implements convolutional 2D filters so that, the weights $W$ of each layer are codified into an array of matrices (a 3D tensor) and produces a vector of processed images (a 3D tensor). On the other hand, classical auto--encoders vectorise the input image and therefore the pixel's spatial relationships are lost. (3) The input of a decoder layer in U--net is the concatenation of the output tensor of the previous layer and output tensor of the symmetric encoder layer (so--called ``skip links''); see Fig. \ref{fig:Vnet}. The purpose of skip links can be understood in the context of residual--nets \cite{he:resnet16}: they allow to construct the solution using both, coarse and processed data.

As in standard convolutional DNNs, U--net follows the thumb--rule for processing inputs tensors: the encoding layers increase the number of channels (filters) of the input tensor, and reduce its spatial dimension, while the decoding layers shrink the number of channels and extend the spatial dimensions of the processed tensors. Therefore, one improves the computational efficiency by applying a reduced number of filters on tensors with larger spatial dimension and a larger number of filters on spatial small--sized tensors. In the training stage, the filters are optimised for detecting the useful features that allow the U--net to correctly classify the pixels.

\subsection*{2.2  The V--net model}
\label{ssec:vnet}

\begin{figure}[ht]
\centering
%\fbox{
\includegraphics[width=\linewidth]{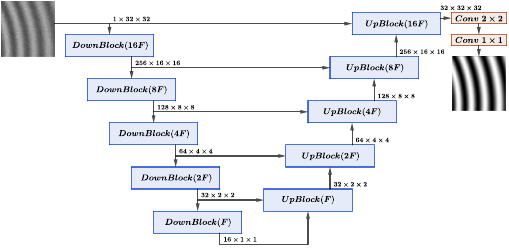}
%}
\caption{The V--net architecture. The number of filters per block is inversely distributed: while U--net is designed for image segmentation (classification), the V--net (shown above) is for image reconstruction (regression).}
\label{fig:Vnet}
\end{figure}

FP normalisation can be understood as a regression problem. Since classification (image segmentation) and regression (image filtering) may require different features, in this work we propose a U--net modification designed to achieve image filtering instead of image segmentation. Despite the computational cost that it implies, our network applies a larger number of filters on the input tensor (original image) in order to capture local details and achieve a precise reconstruction. Also, as opposed to the standard U--net, we reduce the number of filters as the layers are deeper on the encoder. Thus, the deepest layer in the encoder stage (bottom layer on Figure \ref{fig:Vnet}) produces a tensor with the smallest dimension (spatial size and number of channels). These characteristics distinguish our architecture and provide our DNN with an advantage for the regression task. We call our improved model ``V--net'' because it uses tensors with few channels on deeper layers.

According to out knowledge, in the deep learning literature there are no previous studies about the distribution in the number of filters on CNNs. Recently,  we have reported in Refs. \cite{Flores:2020, renteria:2020} that V--net filter distribution out performs the classic U--net filter distribution. In this work, we present the details of outr variant U--net variant, the so--called V--net. So, our approach can be seen as a contribution, that produces better results for the task of FP normalisation and can be used in other image processing task. %at least in the sense of unfollow the common practice of number of filters distribution, in order to produce better reconstructions. 

% ------------------------------------------------------------------------------------------------------------------------------
% ------------------------------------------------------------------------------------------------------------------------------
 
\begin{figure}[htp]
 \centering
\begin{tabular}{cccc}
 \includegraphics[width=\linewidth]{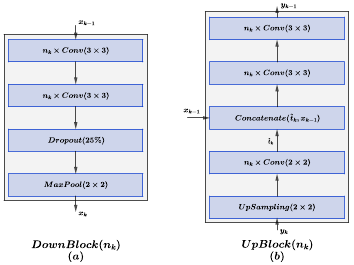} 
 \end{tabular}
 \caption{V--net components: (a) Down--Block and (b) Up--Block.}
	 \label{fig:blocks}
 \end{figure}

The V--net encoder is composed by a sequence of $K$ encoding blocks (Down--Blocks), followed by the decoder, which consist of a sequence of $K$ decoding blocks (Up--Blocks), and a Tail (composed by two last convolutional layers), see Fig. \ref{fig:Vnet} The $k$th Down--Block, $k = 1,2, \ldots, K$, starts by applying two convolutional layers with $n_k$ channels of size $3 \times 3$. This number $n_k$ determines the amount of output channels at each stage. A complete description of each encoding and decoding block architectures is illustrated in Fig. \ref{fig:blocks} and Tables  \ref{tab:downblock} and \ref{tab:upblock}.

In the following, $x_k$ denotes the output tensor of the $k$th Down--Block, $k = 1,2, \ldots, K$, as well as the second input of the ($k+1$)th Up--Block, $k = 1,2, \ldots, K-1$. Similarly, $y_k$ denotes the first input tensor of the $k$th Up--Block, $k = K, \ldots, 2, 1$, as well as the output for the ($k+1$)th Up--Block, $k = K-1, \ldots, 1, 0$.

% ------------------------------------------------------------------------------------------------------------------------------
% ------------------------------------------------------------------------------------------------------------------------------

%-----------------------------------------------------------------------------------------------------------------------------
\begin{table}[h!]
\caption{\bf V--net Down--Block$(a,b,f,p)$ architecture. Parameters: $(None,a,b) =$ shape of input tensor, $f =$ number of filters, $p =$ Dropout rate.}
\centering 
\scalebox{1.0}{
\begin{tabular}{lccc}
\hline
Layer        & Output shape          & \# Params  \\ \hline \hline
Conv2D       & (None, $a$, $b$, $f$) & $3 \times 3 \times f$ \\
Conv2D       & (None, $a$, $b$, $f$) & $3 \times 3 \times f$ \\
Dropout      & (None, $a$, $b$, $f$) & 0       \\
MaxPooling2D & (None, $\lfloor \tfrac{a}{2} \rfloor$, $\lfloor \tfrac{b}{2} \rfloor$, $f$)    & 0       \\ \hline
\end{tabular} }
\label{tab:downblock}
\end{table}
%------------------------------------------------------------------------------------------------------------------------------

%-----------------------------------------------------------------------------------------------------------------------------
\begin{table}[h!]
\caption{\bf V--net Up--Block$(a,b,f)$ architecture. Parameters: $(None,a,b) =$ shape of input tensor, $f =$ number of filters.}
\centering 
\scalebox{1.0}{
\begin{tabular}{lccc}
\hline
Layer        & Output shape             & \# Params  \\ \hline \hline
UpSampling2D & (None, $2a$, $2b$, $f$)  & 0       \\
Conv2D       & (None, $2a$, $2b$, $2f$)  & $2 \times 2 \times f$    \\
Concatenate  & (None, $2a$, $2b$, $4f$) & 0       \\
Conv2D       & (None, $2a$, $2b$, $2f$) & $3 \times 3 \times 2f$   \\
Conv2D       & (None, $2a$, $2b$, $2f$) & $3 \times 3 \times 2f$   \\ \hline
\end{tabular} }
\label{tab:upblock}
\end{table}
%------------------------------------------------------------------------------------------------------------------------------

The complete architecture of our V--net implementation is summarized in Table \ref{tab:vnet}.

%-----------------------------------------------------------------------------------------------------------------------------
\begin{table}[h!]
\caption{\bf V--net architecture.}
\centering 
\scalebox{0.68}{
\begin{tabular}{lcccc}
\hline
\small
Block/Layer    & Output shape        & \# Params & Inputs \\ \hline \hline
01. Input Layer               & (None, 32, 32, 1)   & 0       & \\ \hline
02. DownBlock(32,32,256,0)    & (None, 16, 16, 256) & 592640  & 01     \\
03. DownBlock(16,16,128,0.25) & (None, 8, 8, 128)   & 442624  & 02     \\
04. DownBlock(8,8,64,0.25)    & (None, 4, 4, 64)    & 110720  & 03     \\
05. DownBlock(4,4,32,0.25)    & (None, 2, 2, 32)    & 27712   & 04 \\ \hline
06. UpBlock(2,2,32)           & (None, 4, 4, 32)    & 592640  & 05, 04 \\
07. UpBlock(4,4,64)           & (None, 8, 8, 64)    & 592640  & 06, 03 \\
08. UpBlock(8,8,128)          & (None, 16, 16, 128) & 592640  & 07, 02 \\
09. UpBlock(16,16,256)        & (None, 32, 32, 256) & 592640  & 08, 01 \\ \hline
10. Tail Conv2D       & (None, 32, 32, 2)   & 578     & 09     \\
11. Tail Conv2D       & (None, 32, 32, 1)   & 3       & 10     \\ \hline\hline
Total params         &                 & 3,721,669 & \\
Trainable params     &                 & 3,721,669 & \\
Non-trainable params &                 & 0         & \\ \hline
\end{tabular} }
\label{tab:vnet}
\end{table}
%------------------------------------------------------------------------------------------------------------------------------

In our implementation, we set $K=4$ (so we have four spatial-size levels). The number of channels $n_k$, $k = 1, 2, \ldots, K$, is determined by the number of channels $n_{k-1}$ in the previous Down--Block following the rule $n_k = \tfrac{1}{2} n_{k-1}$. The same occurs with the number of channels for each Up--Block. In our model, the respective number of channels is set as $n_k = 256, 128, 64, 32$, for $k = 1,2, \ldots, 4$. Although more general configurations for the number of channels can be implemented, we have chosen as global parameter the number $F = n_K = 32$ of channels in the last Down--Block (bottom level), and the other $n_k$'s are determined by $F$ as is indicated in Table \ref{tab:vnet}. Figure \ref{fig:Vnet} indicates a deeper version with $K = 5$ blocks and $F = n_5 = 16$. \\

Finally, the training of the V--net can be written as
\begin{equation}
	\label{eq:V--net}
	\argmin{\Theta} \| Y - \hat{Y} \|_1 = \argmin{\Theta} \| Y - (\mathcal T \circ \mathcal D \circ \mathcal E) X_0 \|_1,
\end{equation}
where $X_0$ is the input tensor (stack of all input patches $x_0$), $Y$ is the desired output (stack of all normalised FP patches $y$), and $\hat{Y} = (\mathcal T \circ \mathcal D \circ \mathcal E) X_0$ is the output tensor (stack of all patch estimations $\hat{y}$) of the V--net. The operator $\mathcal T$ represents the Tail stage (last two convolutional layers), added in order to produce a final refinement of the reconstruction. Here, $\Theta$ is the set of all model parameters (filter weights and bias). We use the $L_1$ norm as loss function, because it induces lower reconstruction errors.

%-----------------------------------------------------------------------------------------------------------------------------
\subsection*{2.3  Residual V--net}
\label{ssec:resvnet}
%-----------------------------------------------------------------------------------------------------------------------------

We also propose another two U--net variants in order to explore if there exist variations of our V--net model that even produce better FP normalisations. 

The first variation is a residual version of the V--net. Residual architectures are common in the deep learning architecture design \cite{he:resnet16}, and were introduced to tackle the well--known problem of ``vanishing--gradient'' on deep networks and improve the training process.
The residual learning models can be understood in the context of auto--encoders. While in a typical auto-encoder the encoding-decoding scheme is produced as in equations (\ref{eq:encoder}) and (\ref{eq:decoder}) by $\hat x = (\mathcal{D} \circ \mathcal{E})(x)$, the estimation in residual version of an auto--encoder can be modelled by
\begin{equation}
	\label{eq:encoderdecoder}
	\hat x = x \ominus (\mathcal{D} \circ \mathcal{E})(x) = x \ominus z,
\end{equation}
where $z = (\mathcal{D} \circ \mathcal{E})(x)$.

Here, the original input signal can be understood as a summation
\begin{equation}
	\label{eq:residual}
	x = \hat x \oplus z,
\end{equation}
where the $\hat x$ term refers to the clean (filtered and normalized) signal, plus a noise term $z$, which describes all the perturbation (background, modulation and noise) added to the normalized FP, which is undesired. Hence, the residual scheme learns the noise $z$ that one wants to remove, and can be interpreted as the inverse process to add noise. Residual neural network architectures are common in the image processing community when the associated problem is related to restore and denoise some input signal $x$.

Our implementation of the residual version of the V--net, called here called ``ResV--net", is the following: we have replaced all the Concatenate layers in Table \ref{tab:vnet}, by Subtract layers. That is, instead to concatenate the tensors $\hat t_k$ and $x_{k-1}$ as 
\begin{multline}
\label{eq:concat}
 \tilde t_{k}(m,i,j) = \\ 
	\left\{ 
	\begin{matrix}
		\hat t_{k}(m,i,j)			& m = 1: n_k \\
		x_{k-1} (m-n_k+1,i,j) & m = n_k:n_k+n_{k-1}. 
	\end{matrix}
	\right.
\end{multline}
where $(m,i,j)$ indicate the layer $m$, and the $(x,y)$ spatial position of a input tensor, we replace them by
\begin{equation}
\label{eq:res}
 \tilde t_{k}(m,i,j) = x_{k-1}(m,i,j) \ominus \hat t_{k}(m,i,j), \ m = 1,\ldots,n_k.
\end{equation}

Hence, at each decoder stage (Up blocks) $k = K. \ldots, 2, 1$, the ResV--net model is designed to learn a tensor $\hat t_k$ that removes an amount of the undesired noise from the input signal $x$. Observe that the residual layer introduction reduces the number of channels in each Up--Block from $n_k+n_{k-1}$ to $n_k$, so it reduces the number of parameters in the model. Table \ref{tab:resupblock} describes the redefinition of the Up--Blocks	for the ResV--net. Our implementation of the ResV--net is summarized in Table \ref{tab:resvnet}. \\

A second variation of the proposed V--net is a fast version of the original V--net. It will be described in Section 3.

%-----------------------------------------------------------------------------------------------------------------------------
\begin{table}[h!]
\caption{\bf ResV--net ResUp--Block$(a,b,f)$ architecture. Parameters: $(None,a,b) =$ shape of input tensor, $f =$ number of filters.}
\centering 
\scalebox{1.0}{
\begin{tabular}{lccc}
\hline
Layer        & Output shape             & \# Params  \\ \hline \hline
UpSampling2D & (None, $2a$, $2b$, $f$)  & 0       \\
Conv2D       & (None, $2a$, $2b$, $2f$) & $2 \times 2 \times f$    \\
Subtract     & (None, $2a$, $2b$, $2f$) & 0       \\
Conv2D       & (None, $2a$, $2b$, $2f$) & $3 \times 3 \times 2f$   \\
Conv2D       & (None, $2a$, $2b$, $2f$) & $3 \times 3 \times 2f$   \\ \hline
\end{tabular} }
\label{tab:resupblock}
\end{table}
%------------------------------------------------------------------------------------------------------------------------------

%-----------------------------------------------------------------------------------------------------------------------------
\begin{table}[h!]
\caption{\bf ResV--net architecture.}
\centering 
\scalebox{0.68}{
\begin{tabular}{lcccc}
\hline
\small
Block/Layer    & Output shape        & \# Params & Inputs \\ \hline \hline
01. Input Layer               & (None, 32, 32, 1)   & 0       & \\ \hline
02. DownBlock(32,32,256,0)    & (None, 16, 16, 256) & 592640  & 01     \\
03. DownBlock(16,16,128,0.25) & (None, 8, 8, 128)   & 442624  & 02     \\
04. DownBlock(8,8,64,0.25)    & (None, 4, 4, 64)    & 110720  & 03     \\
05. DownBlock(4,4,32,0.25)    & (None, 2, 2, 16)    & 27712   & 04 \\ \hline
06. ResUpBlock(2,2,32)        & (None, 4, 4, 32)    & 592640  & 05, 04 \\
07. ResUpBlock(4,4,64)        & (None, 8, 8, 64)    & 592640  & 06, 03 \\
08. ResUpBlock(8,8,128)       & (None, 16, 16, 128) & 592640  & 07, 02 \\
09. ResUpBlock(16,16,256)     & (None, 32, 32, 256) & 592640  & 08, 01 \\ \hline
10. Tail Conv2D       & (None, 32, 32, 2)   & 578     & 09     \\
11. Tail Conv2D       & (None, 32, 32, 1)   & 3       & 10     \\ \hline\hline
Total params         &                 & 3,721,669 & \\
Trainable params     &                 & 3,721,669 & \\
Non-trainable params &                 & 0         & \\ \hline
\end{tabular} }
\label{tab:resvnet}
\end{table}
%------------------------------------------------------------------------------------------------------------------------------

% ------------------------------------------------------------------------------------------------------------------------------
\section*{3.  Implementation details}
\label{sec:details}
% ------------------------------------------------------------------------------------------------------------------------------

%-----------------------------------------------------------------------------------------------------------------------------
\subsection*{3.1  Simulated data}
\label{ssec:simulated}
%-----------------------------------------------------------------------------------------------------------------------------

To quantitatively evaluate the performance of the proposed V--net based normalisation, we randomly two datasets: the first consist of 46 generated pairs of FPs with size $1024 \times 1024$ pixels; the second one consist of 180 generated FPs with size $320 \times 320$ pixels. In both cases, the corrupted FPs were generated according to the model in \eqref{eq:fp} and the normalised FPs (ground--truth) according to the model in \eqref{eq:fpnorm}. The normally distributed random noise was generated with the Python Numpy package. On the other hand, the random smooth functions (illumination components and phase) were constructed using random numbers with uniform distribution generated with our implementation of a Linear Congruential Generator with POSIX parameters \cite{posix:rotenberg60} in order to guarantee the FPs generation replicability. In the following, we explain the smooth random surface generation procedure.

\begin{figure}[t!]
\centering
%\fbox{
\includegraphics[width=.8\linewidth]{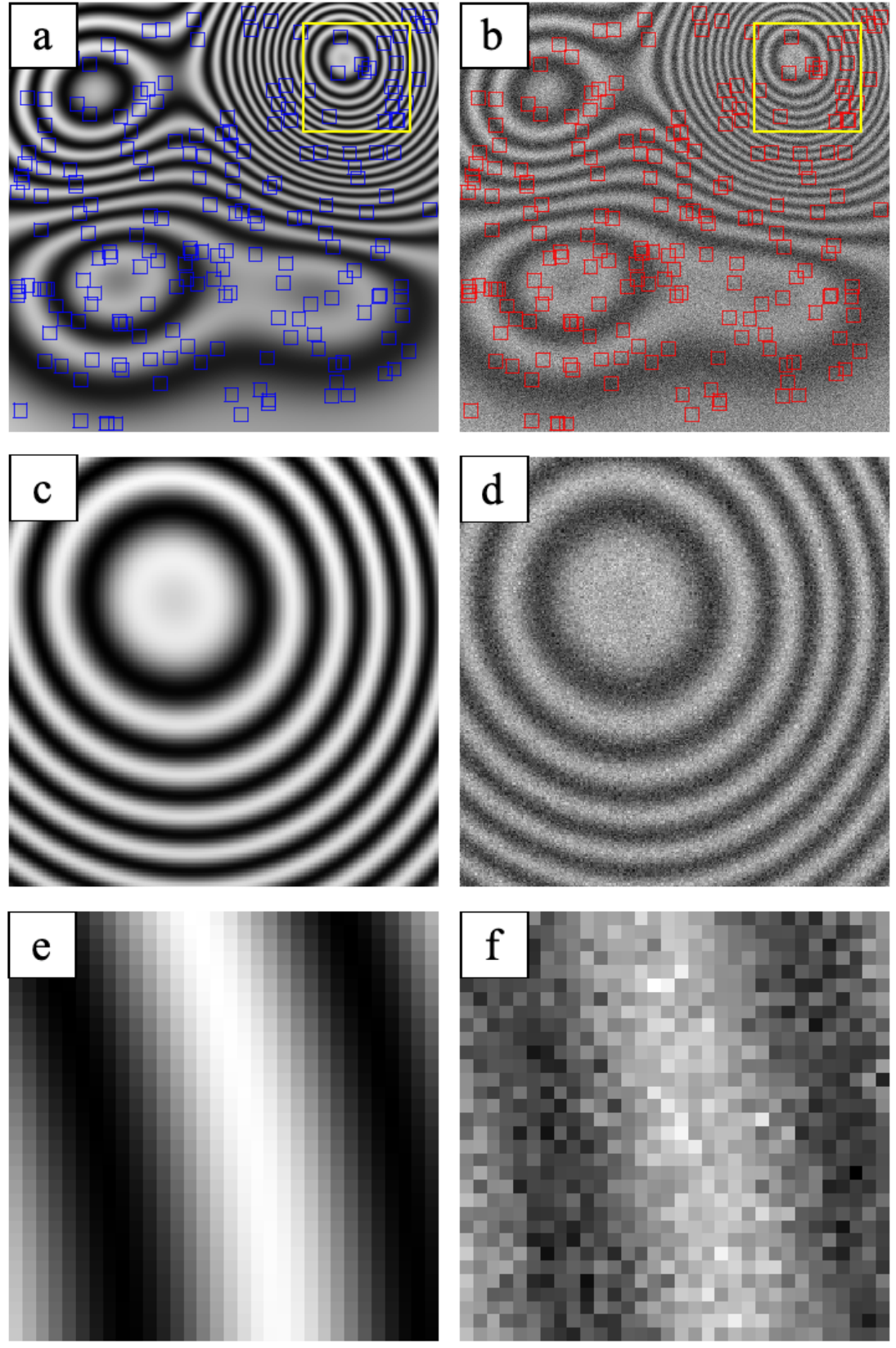}
%}
\caption{Example of training data. (a) Synthetic normalised FP of $1024 \times 1024$ pixels with a selected region of interest of $256 \times 256$ pixels (yellow square), small random patches of $32 \times 32$ in blue; (b) the same FP corrupted with Gaussian noise, patches in red; (c) and (d) regions of interest in (a) and (b), respectively. A random patch--pair used for training: (e) Ground--truth and (f) corrupted input.}
\label{fig:train_data}
\end{figure}

We generated the pseudo--random phase with a radial basis function with Gaussian kernel \cite{rbf:broomhead88}:

\begin{equation}
\label{eq:sint_phi}
\phi(p) = \sum_{i=1}^{10} \alpha_i G(p; m_i, \sigma_{\phi}),
\end{equation}
where we define
\begin{equation}
\label{eq:G}
G(p; \mu, \sigma) \overset{def}{=} \exp \left(-\frac{1}{2 \sigma^2} \| p - \mu \|^2 \right),
\end{equation}
$m = \left[ m_1, m_2, \ldots, m_{10} \right]^\top$ the vector of the random kernels centers that are uniformly distributed into the FPs lattice (\ie, $m_i \in [0,1023]^2$), and $\alpha$ the vector of random uniformly--distributed Gaussian heights, with $\alpha_i \in [-180/\pi, 180/\pi]$. Similarly, we generated the illumination term with $a(p) = G(p, m_a; \sigma_a) $ and $b(p) = G(p, m_b; \sigma_b)$. In our data, we selected $m_a$ and $m_b$ uniformly distributed over the image domain (using our implementation of the POSIX algorithm) and we set $\sigma_{\phi} = 1024/6$, $\sigma_a = 1024/2$ and $\sigma_b = 1024$. Figure \ref{fig:train_data} depicts an example of the synthetic data used for training: Panels \ref{fig:train_data}(a) and \ref{fig:train_data}(b) show the ideal and corrupted FPs, respectively. Panels \ref{fig:train_data}(c) and \ref{fig:train_data}(d) show a selected region. An example of a patch-pair of size $32 \times 32$ used for training is depicted in panels \ref{fig:train_data}(e) and \ref{fig:train_data}(f). \\

In the Experiments section, we evaluate our method performance for different noise types, as Gaussian, salt--pepper, speckle, and combinations of them. We have also evaluated the performance of the models when some circular field of vision restriction mask (FOV), is added. 

For the salt--pepper noise experiment, we randomly select the $25 \%$ of the pixels and saturate them to values $0$ and $1$ in equal proportion. In the case of FPs with speckle noise (ESPI), we generate the fringes according to the model
\begin{align}
 	\label{eq:spckl}
	x(p) = & a(p) +  b(p) | \cos\left( \phi(p) + \eta_1(p) \right) + \cos \eta_1(p) | \nonumber \\
	           & + \eta_2(p)
\end{align}
 instead of \eqref{eq:fp}; where $\eta_1$ and $\eta_2$ are spatially independent and identically distributed noise: $\eta_1$ has uniform distribution (with values into $[1,100]$ radians) and $\eta_2$ has Gaussian distribution (with zero mean and standard deviation $\sigma_\eta = 2.5$).

% ------------------------------------------------------------------------------------------------------------------------------
% ------------------------------------------------------------------------------------------------------------------------------

% ------------------------------------------------------------------------------------------------------------------------------
\begin{figure}[ht]
\centering
%\fbox{
\includegraphics[width=.8\linewidth]{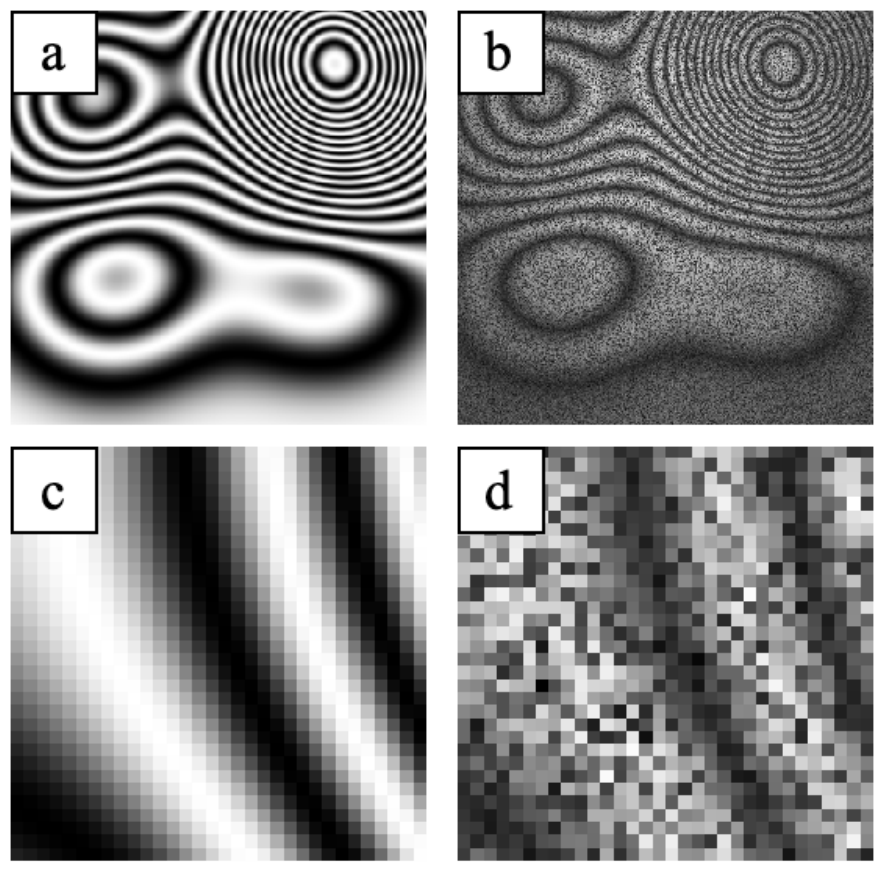}
%}
\caption{Example of high-noised training data. (a) Synthetic normalised FP of $320 \times 320$ pixels; (b) the same FP corrupted with ESPI noise. A random patch--pair of $32 \times 32$ used for training: (c) Ground--truth and (d) corrupted input.}
\label{fig:noisy_data}
\end{figure}
% ------------------------------------------------------------------------------------------------------------------------------

% ------------------------------------------------------------------------------------------------------------------------------
\begin{figure}[ht]
\centering
%\fbox{
\includegraphics[width=.8\linewidth]{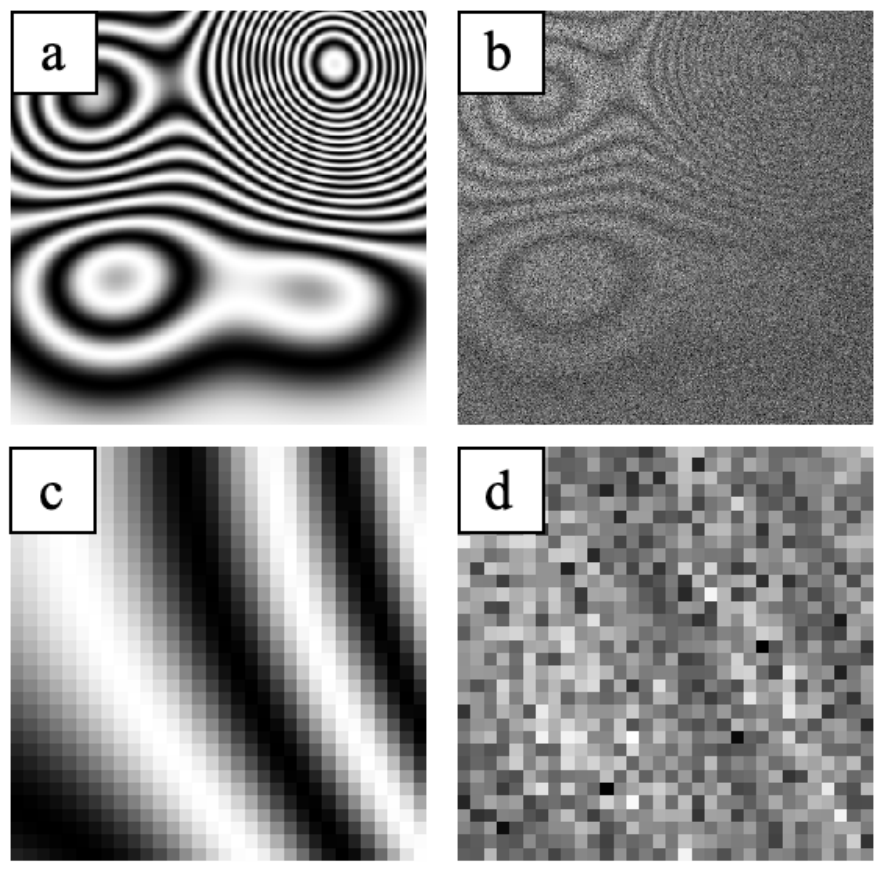}
%}
\caption{Example of extreme-noise training data. (a) Synthetic normalised FP of $320 \times 320$ pixels; (b) the same FP corrupted with ESPI and gaussian noise. A random patch--pair of $32 \times 32$ used for training: (c) Ground--truth and (d) corrupted input.}
\label{fig:extreme_data}
\end{figure}
% ------------------------------------------------------------------------------------------------------------------------------

In the ESPI case, we have also evaluated two more testing experiments with extremal noise conditions. For those experiments, we have considered the second dataset of FPs with size $320 \times 320$. This FP dataset is generated according to model (\ref{eq:spckl}), in two scenarios:
\begin{itemize}
	\item[i)] \underline{High noise}: $\eta_1$ has uniform distribution in $[0,1000]$ (radians) and $\eta_2$ is a low Gaussian distribution (with zero mean and $\sigma_\eta = 2.5$, and modulation amplitude $b(p)$ generated as a Gaussian with maximal amplitude $b = 0.4$ (equivalent to to 40\% of the signal). See Figure \ref{fig:noisy_data}.
	\item[ii)] \underline{Extreme noise}: $\eta_1$ has uniform distribution in $[0,1000]$ (radians) and $\eta_2$ is a high Gaussian distribution (with zero mean and $\sigma_\eta = 2.5$, and modulation amplitude $b(p)$ generated as a gaussian with maximal amplitude $b = 1.25$ (equivalent to to 125\% of the signal). See Figure \ref{fig:extreme_data}.
\end{itemize}

%-----------------------------------------------------------------------------------------------------------------------------
\subsection*{3.2  Training data sets}
\label{ssec:dataset}
%-----------------------------------------------------------------------------------------------------------------------------

The training data generated from first image dataset consists of $25,000$ random patches of $32 \times 32$ pixels sampled from the first 30 generated FPs (the set of training images); $2500$ of those patches were used for validation. In addition, the remaining $16$ last FPs were used as the test data set, in order to measure and evaluate the performance of all compared models. We stacked the corrupted patches in the tensor $X_0=[ x_i ]_{1=1,2,\ldots, 25,000}$ and the corresponding normalised patches form the desired output $Y = [ y_i ]_{1=1,2,\ldots, 25,000}$. The input data for all our models is the pair of tensors $X_0$ and $Y$.

For the highly-noised and extreme-noised experiments. The training data was generated from the seconda image dataset. This consists of $40,000$ random patches of $32 \times 32$ pixels sampled from the first 150 generated FPs (the set of training images); $4000$ of those patches were used for validation. In this case, the remaining $30$ last FPs were used as the test data set, for evaluation purposes. Similar to the previous experiment, the tensors $X_0=[ x_i ]_{1=1,2,\ldots, 40,000}$ and the corresponding normalised patches $Y = [ y_i ]_{1=1,2,\ldots, 40,000}$ form the input data for all tested models.

In both cases, the patch--size of 32 is a user-defined parameter. We chose the size as $32 \times 32$ by considering a maximum frequency close to 1.5 fringes per patches. Moreover, the V--net also requires a patch--size divisible by $2^K = 2^5 = 32$; where $K$ is the number of levels or blocks included in the model design.

%-----------------------------------------------------------------------------------------------------------------------------
\subsection*{3.3  Prediction of a full FP from reconstructed patches}
\label{ssec:full_fp}
%-----------------------------------------------------------------------------------------------------------------------------

\begin{figure}[t]
\centering
%\fbox{
\includegraphics[width=0.9\linewidth]{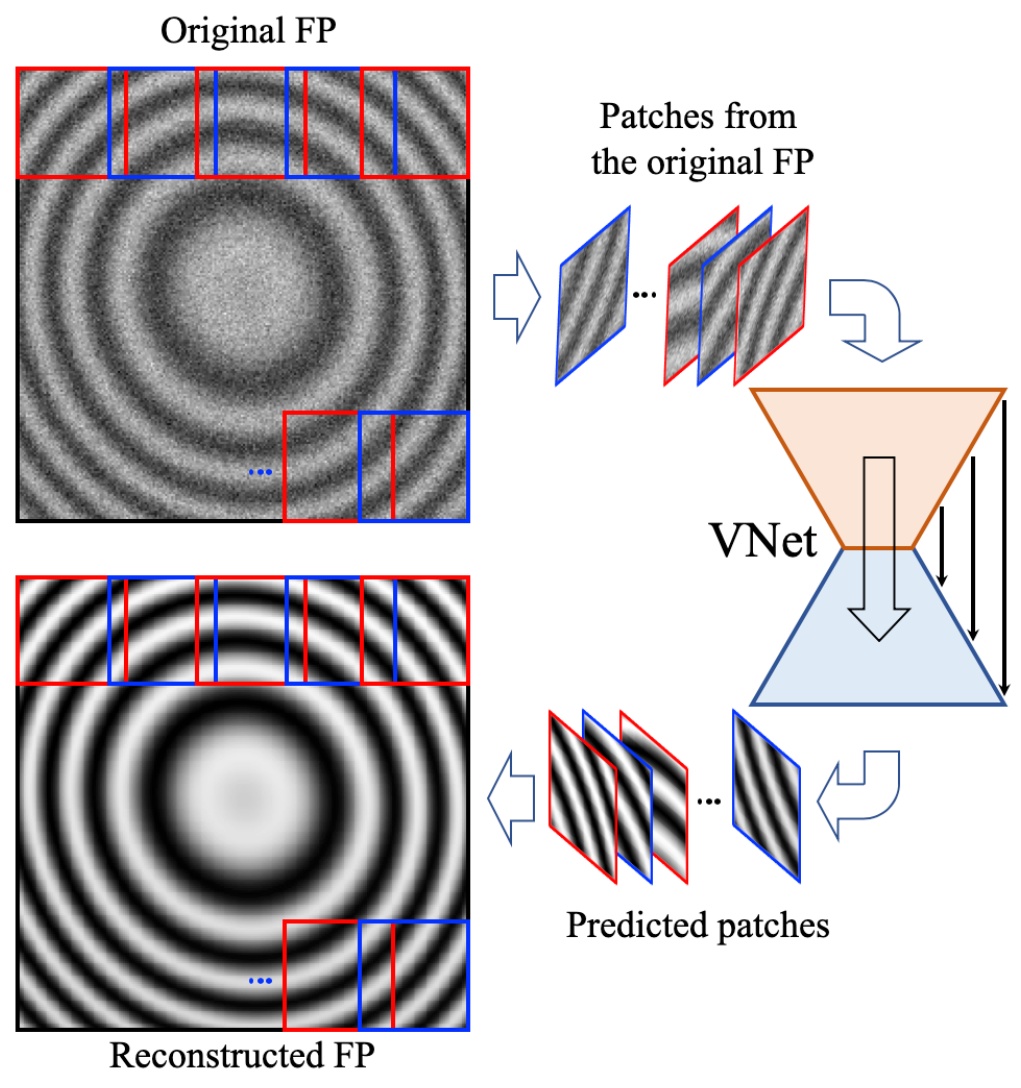}
%}
\caption{FP normalisation (inference). A set of overlapped patches that cover the entire FP to normalise is computed and used to feed our trained V--net model, the predicted patches are assembled to reconstruct the FP original. Pixels with multiple predictions (because of the patches' overlapping) are averaged for computing the normalised (reconstructed) FP.}
\label{fig:inference}
\end{figure}

Recall that the V--net is designed to reconstruct small FP patches of $32 \times 32$ pixels. Thus, to reconstruct an entire FP, we generated a set of patches using a sliding window scheme, with a stride (pixels shifts step) of $s_x = s_y = 4$ pixels in both horizontal and vertical directions in the $1024 \times 1024$ images, while $s_x = s_y = 2$ for the $32 \times 32$ datasets. All patches are fed to the V--net to compute their normalisations; see Fig. \ref{fig:inference}. Each pixel in the entire reconstructed FP was computed as the average of the values in the same pixel position obtained from overlapped normalised patches. We preferred the mean because it is more efficiently computed than the median, and we did not appreciate a significative difference if the median is used instead.

Again, the pixel shifts $s_x$ and $s_y$ are user-defined parameters. We have also tested pixel shifts equal to 2 and 1, but the improvement in the recostruction is not significant, and the selection of lesser values of $s_x$ and $s_y$ only increases the computational cost and time (nearly $\times 4$ times for the 2 stride, and $\times 16$ times in che case of the 1 stride). Higher values of $s_x, s_y$ reduce the computational cost but can produce bad non-smooth reconstructions by introducing an undesired checker effect. Therefore, our choice of $s_x = s_y = 4$ responds to a trade-off between good reconstructions and computational cost.

% ------------------------------------------------------------------------------------------------------------------------------
% ------------------------------------------------------------------------------------------------------------------------------

For a 2-dimensional FP image, let $\kappa_d$ be the number of patches computed over the dimensions $d = 1,2$ (rows and columns). Then, $\kappa_d$ is given by
\begin{equation}
	\kappa_d = q_d + \varepsilon_d;
\end{equation}
where 
\begin{equation}
	q_d = \left \lfloor \frac{H_d - h_d }{s_d} \right \rfloor + 1,
\end{equation}
$H_d$ is the image size, $h_d$ the patch size, $s_d$ the stride step and
\begin{equation}
	\varepsilon_d = \left\{ \begin{matrix}
	1 & H_d - (q_d-1) s_d - h_d >0 \\
	0 & \text{otherwise}.
	\end{matrix} \right.
\end{equation}

In the case of the $1024 \times 1024$ dataset, we set $H_d=1024$, $h_d=32$, $s_d=4$ for $d=1,2$. Then, the number of patches required to reconstruct a single FP is $\kappa_1\times \kappa_2 = 249 \times 249 = 62,001$. This quantity is substantially larger than the number of patches in the training set, $25,000$ patches. The expected number of patches per training FP was 833 ($25,000/30)$. We used Montecarlo simulations to estimate the covered area by the selected patches: in average, it was $53\%$ of each entire FP. If we modify the number of training patches to $40,000$, the averaged covered area would be $71\%$ of each FP.

%-----------------------------------------------------------------------------------------------------------------------------
\subsection*{3.4  The fast V-net implementation}
\label{ssec:fastvnet}
%-----------------------------------------------------------------------------------------------------------------------------

Finally, a second modification of our V--net implementation is proposed. This variant preserves the same architecture of the V--net described in Figure \ref{fig:Vnet} and Table \ref{tab:vnet}, just modifies the shape of the input tensor. 

The modification is as follows: 1) After the V--net is trained, we directly clone our V--net model, with all hyper-parameters maintained the same, and all learned filters (or weight parameters) remain equal. We just duplicate our model, now setting the input tensor shape, as equal as the FP image dimensions ($1024 \times 1024$ or $320 \times 320$, depending the experiment scenario). 2) Now we transfer all weigths, layer by layer, from the original V--net to this new clone. The obtained clone version is what we call here the ``Fast V--net". \\

Now the reconstruction or inference process is done by passing the entire corrupted FP to the network, instead of the set of $32 \times 32$ covering patches. Thas is, we generate the normalized FP using one-patch only of the same size of the image: the entire FP. As a result, we obtain similar quality of reconstructions, but now the computational time needed to process a single FP reduces to a fraction of the time consumed by the V--net. The reduction is about $100$ times faster that the original V--net implementation. Table \ref{tab:fastvnet} summarizes the architecture of the Fast V--net. It is identical to the architecture in \ref{tab:vnet}, only that all output shapes are rescaled according to the input shape (\eg $\times 10$ in the case of a $320 \times 320$ input image).

% ------------------------------------------------------------------------------------------------------------------------------
\begin{figure}[hb]
\centering
%\fbox{
\includegraphics[width=0.7\linewidth]{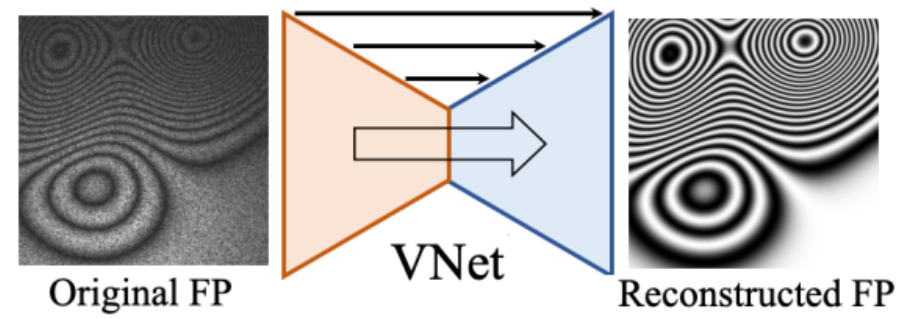}
%}
\caption{1-patch Fast V--net normalisation (inference).}
\label{fig:fastinference}
\end{figure}
% ------------------------------------------------------------------------------------------------------------------------------

%-----------------------------------------------------------------------------------------------------------------------------
\begin{table}[h!]
\caption{\bf Fast V--net architecture (Just for inference, images of size $320 \times 320$).}
\centering 
\scalebox{0.66}{
\begin{tabular}{lcccc}
\hline
\small
Block/Layer    & Output shape        & \# Params & Inputs \\ \hline \hline
01. Input Layer               & (None,320,320,1)   & 0       & \\ \hline
02. DownBlock(320,320,256,0)    & (None,160,160,256) & 592640  & 01     \\
03. DownBlock(160,160,128,0.25) & (None,80,80,128)   & 442624  & 02     \\
04. DownBlock(80,80,64,0.25)    & (None,40,40,64)    & 110720  & 03     \\
05. DownBlock(40,40,32,0.25)    & (None,20,20,32)    & 27712   & 04 \\ \hline
06. UpBlock(20,20,32)           & (None,40,40,32)    & 592640  & 05, 04 \\
07. UpBlock(40,40,64)           & (None,80,80,64)    & 592640  & 06, 03 \\
08. UpBlock(80,80,128)          & (None,160,160,128) & 592640  & 07, 02 \\
09. UpBlock(160,160,256)        & (None,320,320,256) & 592640  & 08, 01 \\ \hline
10. Tail Conv2D       & (None,320,320,2)   & 578     & 09     \\
11. Tail Conv2D       & (None,320,320,1)   & 3       & 10     \\ \hline\hline
Total params         &                 & 3,721,669 & \\
Trainable params     &                 & 3,721,669 & \\
Non-trainable params &                 & 0         & \\ \hline
\end{tabular} }
\label{tab:fastvnet}
\end{table}
%------------------------------------------------------------------------------------------------------------------------------

%-----------------------------------------------------------------------------------------------------------------------------
\section*{4.  Experiments}
\label{sec:experiments}
%-----------------------------------------------------------------------------------------------------------------------------

In order to evaluate the performance of the U--net and the proposed V--net models for the FP normalisation task, we conducted three experiments. In the first one, we evaluated the U--net and V--net performance with respect to the noise level (assuming Gaussian noise). In the second experiment, we evaluated such models under different noise distributions: Gaussian, salt--pepper, speckle, combination of noise and the effect of incomplete field of view (named Pupil in this work). Finally, the third experiment compares our proposals with methods of the state of the art, in normal scenarios and when  high-noise and extreme noise is added to the FPs.

For all the evaluated networks, we equally set parameters for the training process. We used the ADAM algorithm \cite{adam:kingma15} as optimiser with a learning rate $1\times10^{-4}$, a decay rate $1\times10^{-3}$, a batch size equal to 32 and we select the best trained model over 300 epochs.

%-----------------------------------------------------------------------------------------------------------------------------
\begin{table}[tp]
\centering
\caption{\bf Summary of the synthetic experiments (full-images). FPs were generated using \eqref{eq:fp}.}
\begin{tabular}{ccc}
\hline
Standard deviation & U--net MAE & V--net MAE \\
 ($\sigma$ with $a,b$ variable) & ($\times 10^{-4}$) & ($\times 10^{-4}$) \\
 \hline 
0.00			&	{\bf 2.142}	&	2.266	\\
0.05			&	{\bf 2.016}	&	2.383	\\
0.10			&	{\bf 2.416}	&	2.457	\\
0.15			&	2.552	&	{\bf 2.539}	\\
0.20			&	2.762	&	{\bf 2.620}	\\
0.25			&	2.769	&	{\bf 2.726}	\\
0.30			&	2.807	& {\bf 2.702}	\\
 \hline
\end{tabular}
\label{tab:exps_gss}
\end{table}
%------------------------------------------------------------------------------------------------------------------------------

\subsection*{4.1  Performance comparison of U--net and V--net for different noise levels}
\label{ssec:noise_level}

In this experiment, we simulated noise levels as in the acquisition of typical interferometric FPs. We investigated the performance of the U--net and V--net models for seven levels of Gaussian noise; \ie, seven standard deviations $\sigma$ for the noise $\eta$ in \eqref{eq:fp}. Such $\sigma$ values are indicated in first column in Table \ref{tab:exps_gss}. For each trained model, we used a randomly generated training set and a randomly generated initial starting point for the models' parameters (weights). Table \ref{tab:exps_gss} reports the averaged Mean--Absolute--Error (MAE) of the reconstructions over ten different trained models. 

\begin{figure}[ht]
 \centering
 \includegraphics[width=\linewidth]{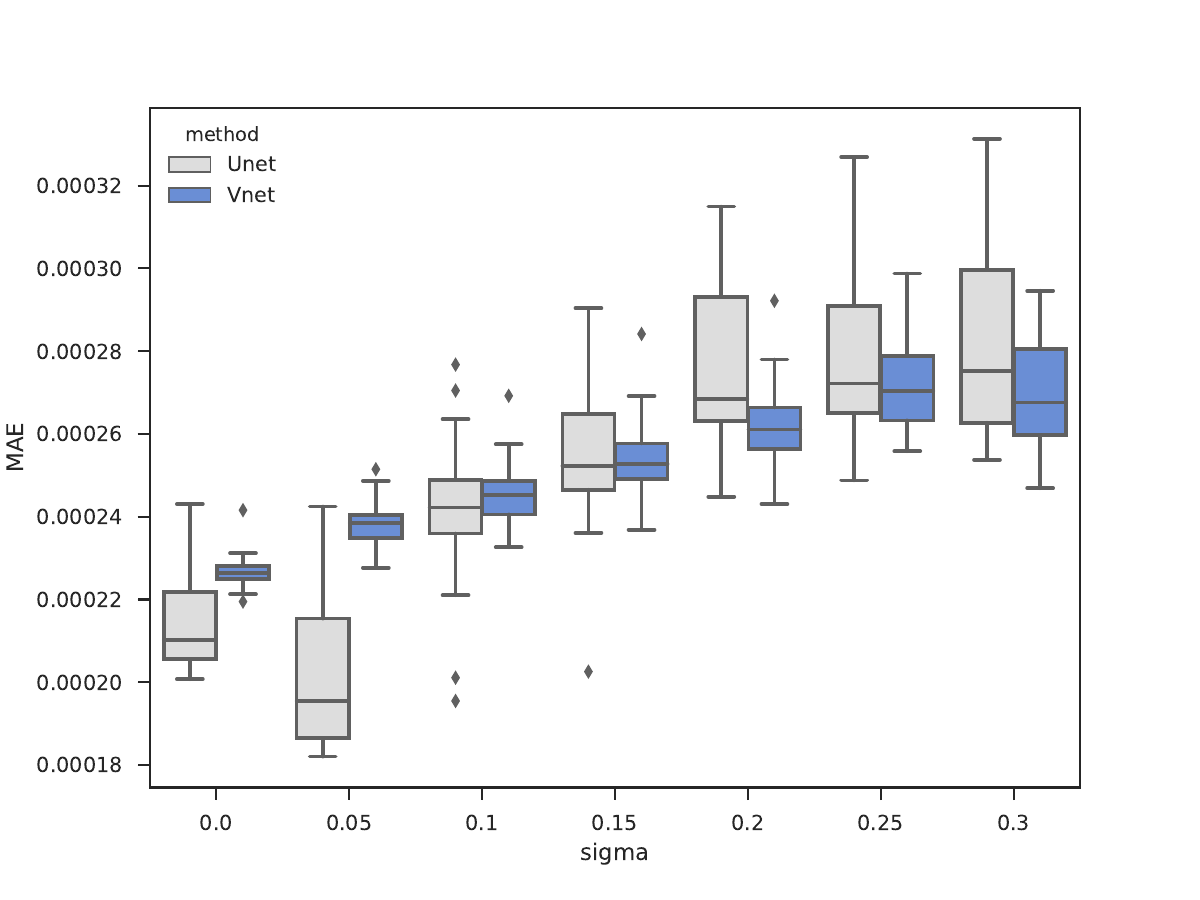} 
 \caption{Summary of experiments for normalising 46 FPs corrupted with Gaussian noise and different levels of $\sigma$.}
	\label{fig:exps_gss}
 \end{figure}

\begin{figure}[h!]
 \centering \includegraphics[width=0.95\linewidth]{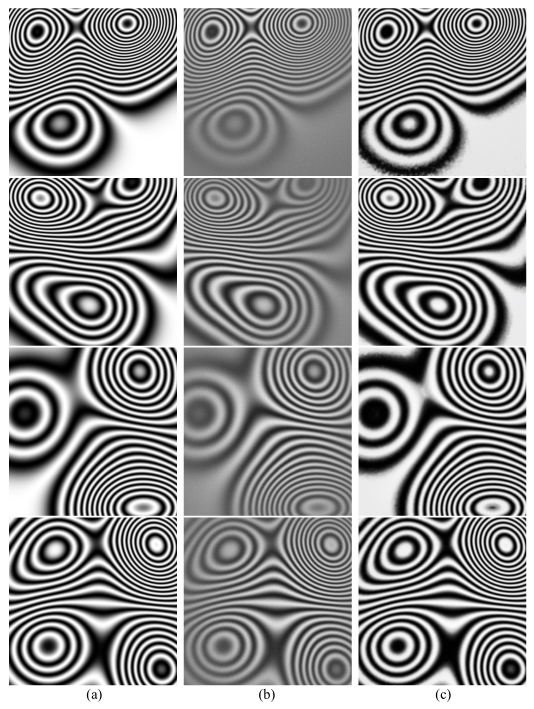}
 \caption{Denoised--normalised FPs with V--net: a) Ground--truth; b) corrupted FPs, $x$; and c) reconstructions, $\hat y$.}
	\label{fig:FPs}
 \end{figure}

Table \ref{tab:exps_gss} shows that, in general, V--net performs better than U--net for denoising FPs corrupted with Gaussian noise. According to Fig. \ref{fig:exps_gss}, the V--net model has a superior performance for higher standard deviation values ($\sigma > 0.15$ with a signal's dynamic range into the interval $[0,2]$). Both models have a similar performance for $\sigma$ close to $0.1$. On the other hand, U--net produces better reconstructions for low noise levels ($\sigma \le 0.05$). Fig. \ref{fig:FPs} shows examples of FPs normalised with our method (noise with $\sigma = 0.15$). In general, V--net presents lower error variance, that is understood as a better precision of the results.

% ------------------------------------------------------------------------------------------------------------------------------
% ------------------------------------------------------------------------------------------------------------------------------

% ------------------------------------------------------------------------------------------------------------------------------

% ------------------------------------------------------------------------------------------------------------------------------

%------------------------------------------------------------------------------------------------------------------------------
\subsection*{4.2  Performance comparison of U--net and V--net for different noise types}
\label{ssec:noise_types}
%------------------------------------------------------------------------------------------------------------------------------

The following experiment reports the U--net and V--net model's performance for the normalisation of FPs under different noise distributions; in all cases, the illumination components $(a,b)$ and the phase $\phi$ were generated according to the method presented in subsection \ref{ssec:simulated}. The pupil was defined with a centered circular region of diameter equal to $80\%$ of the image size.

% ------------------------------------------------------------------------------------------------------------------------------

\begin{table}[t!]
\centering
\caption{\bf Summary of the synthetic experiments (patches). Speckle FPs were generated using \eqref{eq:spckl}.}
\scalebox{0.92}{
\begin{tabular}{lcc}
\hline
Noise & U--net MAE & V--net MAE \\
 ($\eta$ with $a,b$ variable) & ($\times 10^{-4}$) & ($\times 10^{-4}$) \\
 \hline
$\eta=0$ (no noise)			&	{\bf 2.142}	&				2.266	\\
Salt-pepper							&	{\bf 2.416}	&				2.455	\\
Speckle									&				2.552	&	{\bf 2.539}	\\
Gaussian-speckle				&				2.762	&	{\bf 2.620}	\\
Gaussian-speckle-pupil	&				2.769	&	{\bf 2.726}	\\
\hline
\end{tabular} }
	\label{tab:exps_types}
\end{table}

% ------------------------------------------------------------------------------------------------------------------------------
\begin{figure}[t!]
 \centering
 \includegraphics[width=\linewidth]{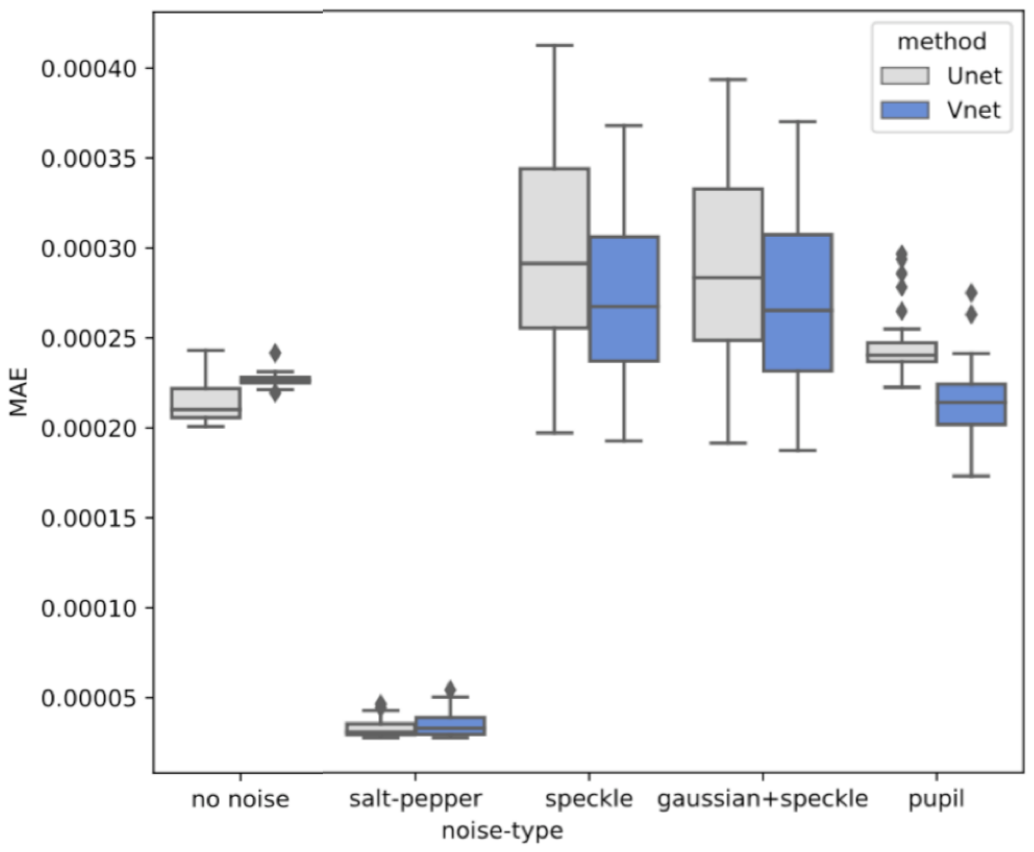} 
 \caption{Summary of experiments for normalising 46 FPs corrupted with different noise distributions.}
	\label{fig:exps_types}
 \end{figure}
% ------------------------------------------------------------------------------------------------------------------------------

Table \ref{tab:exps_types} and Fig. \ref{fig:exps_types} report results for corrupted FPs under different scenarios: Salt--pepper noise, Speckle noise, Gaussian--and--speckle noise and an incomplete field of view (pupil). Note that V--net produces better results for Speckle noise, Gaussian+Speckle noise and pupil. In contrast, U--net has a better performance when the task requires processing data with few intensity levels: as in the reconstruction of FP corrupted with salt--pepper noise (to remove a few data and to interpolate such pixels) or if only low--frequency illumination changes are present and the noise is not a problem.

%------------------------------------------------------------------------------------------------------------------------
\subsection*{4.3  Comparison versus state of the art methods}
\label{ssec:sotam}
%------------------------------------------------------------------------------------------------------------------------

In this subsection we evaluate the performance of the proposed models U--net, V--net, ResV--net versus other methods of state of the art based on deep neural networks.

We compared our proposed models, with recently reported Deep Neural Networks: Optical Fringe Patterns Denoising (FPD) convolutional neural network proposed in Ref. \cite{fpd:lin19}, Deep Convolutional Neural Network (DCNN) \cite{dcnn:yan19} and the application reported in Ref. \cite{ffd:hao19} of the general purpose image denoising deep neural network (FFD) \cite{ffdnet:zhang18}. 

\begin{figure}[ht]
 \centering
 \includegraphics[width=0.95\linewidth]{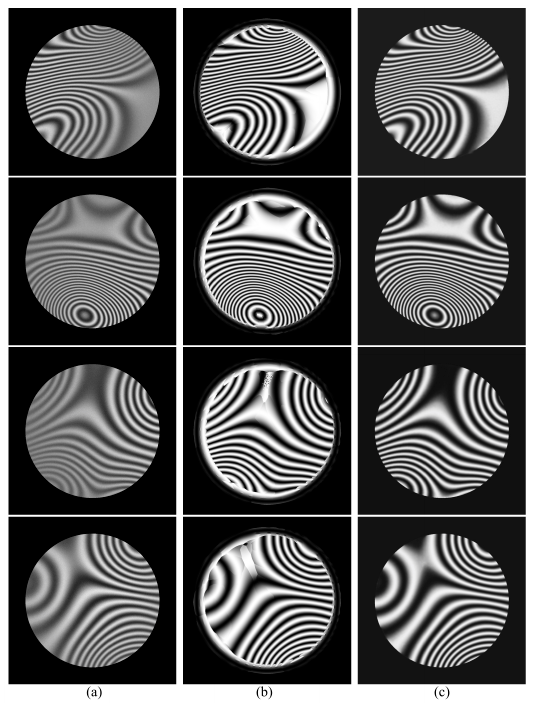} 
 \caption{Normalised FPs. (a) Data generated with Gaussian and speckle noise. (b) GFB based normalisation. (c) Our results.}
 \label{fig:gfb}
 \end{figure}

In Refs. \cite{dcnn:yan19, fpd:lin19} are presented favourable comparisons of their networks with respect to a filtering based on the Windowed Fourier Transform (WFT) \cite{huang:waveletFP10, Kemao:04}. The authors argue they have chosen WFT since it is one of the classical procedures with better performance for fringe denoising. We have compared our proposals with a particular case of WFT: the Gabor Filter Bank---in Ref. \cite{rivera:transient18} is reported the relationship between GFB and WFT. The results of our comparisons are consistent with the reported in Ref. \cite{dcnn:yan19}: the GFB method fails to reconstruct the FP at regions with phase discontinuities and low--frequency. Figure \ref{fig:gfb} depicts evidence that supports this claim.

% ------------------------------------------------------------------------------------------------------------------------------
% ------------------------------------------------------------------------------------------------------------------------------

% ------------------------------------------------------------------------------------------------------------------------------

\begin{figure*}[htp]
 \centering \centering \includegraphics[width=\linewidth]{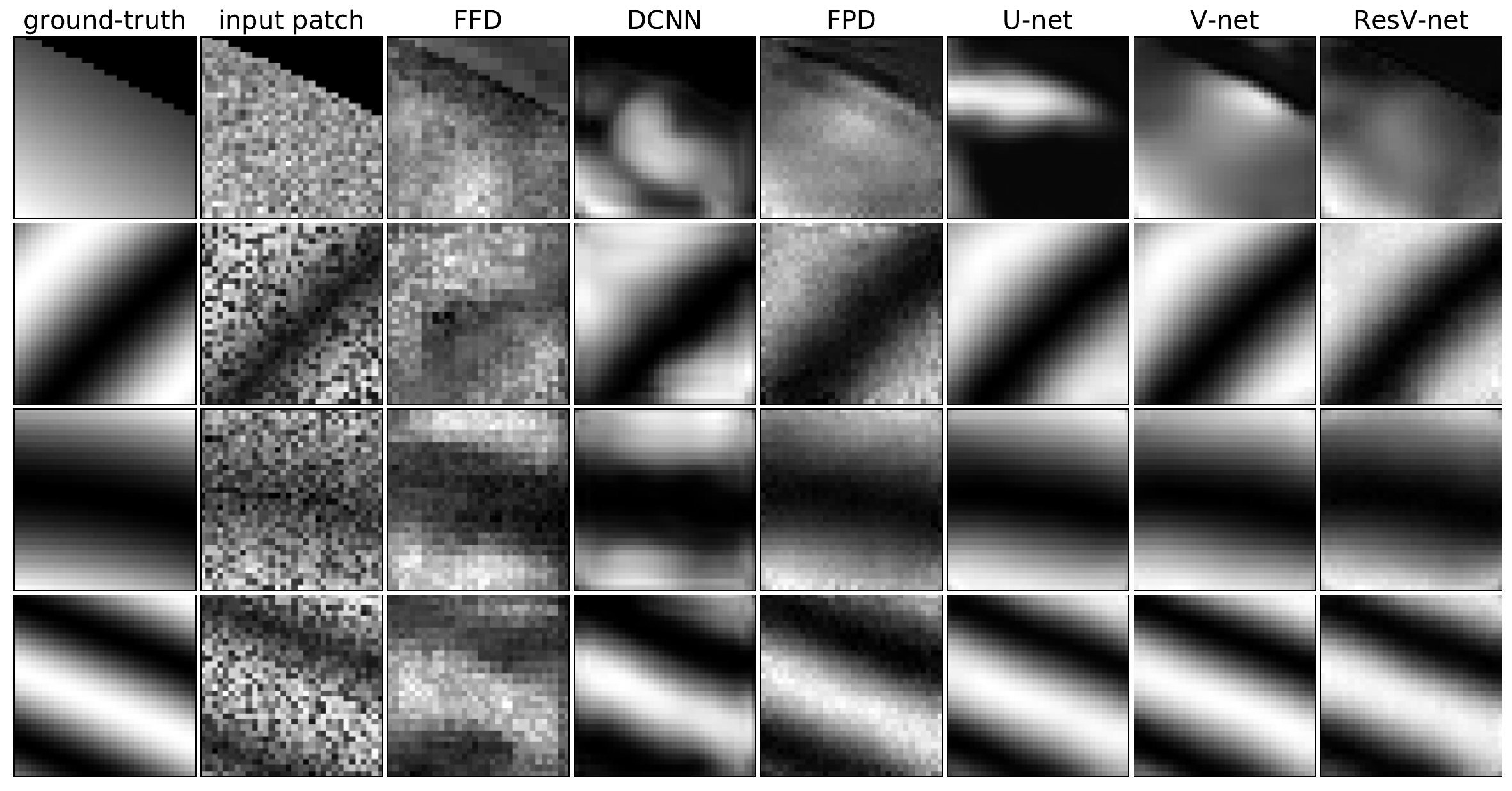}
 \caption{Results of the compared deep neural networks models: normalised patches.}
	 \label{fig:nets_patches}
 \end{figure*}
% ------------------------------------------------------------------------------------------------------------------------------

% ------------------------------------------------------------------------------------------------------------------------------

\begin{figure*}[htp]
 \centering \centering \includegraphics[width=\linewidth]{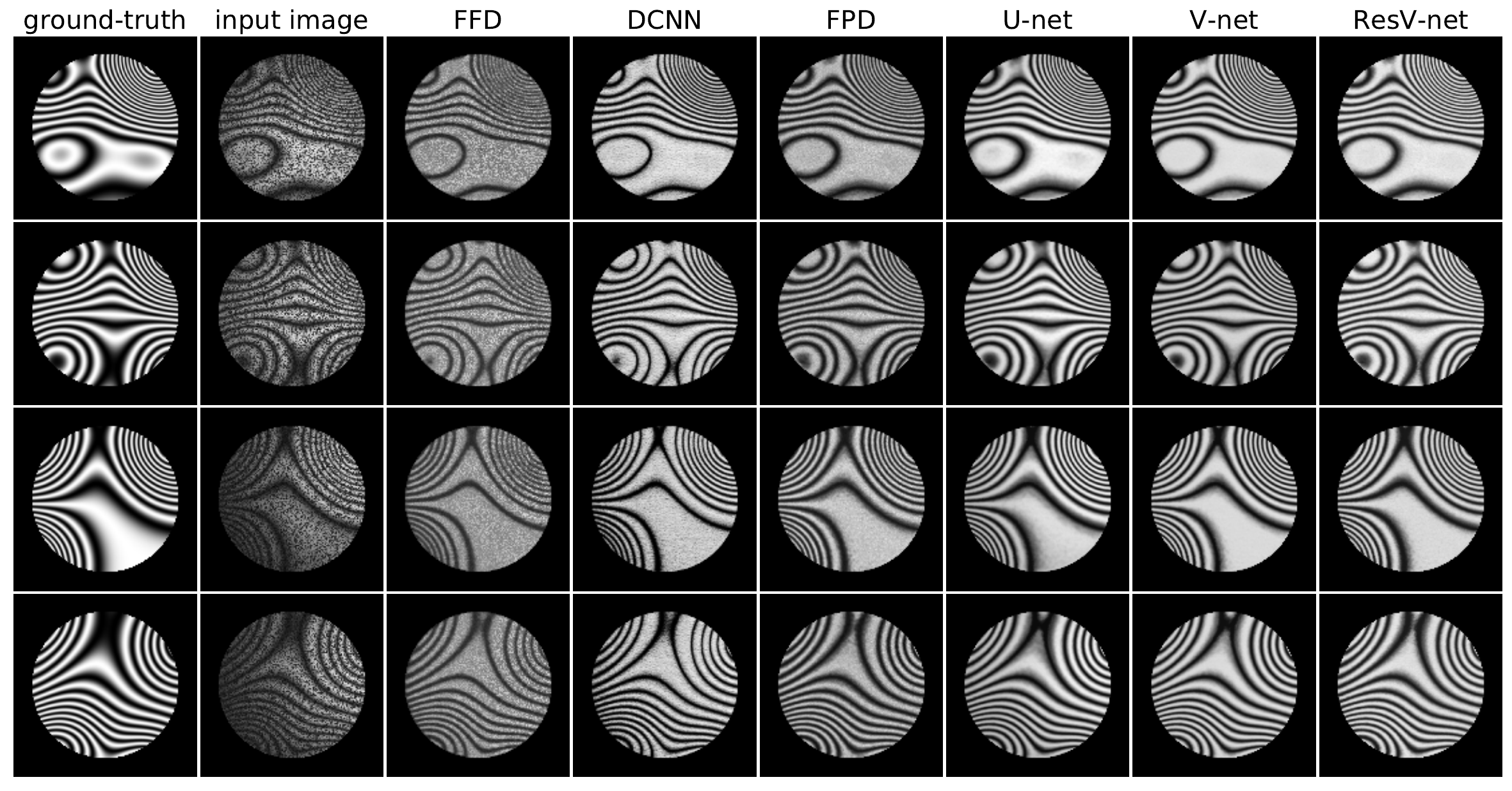}
	\caption{Results of the compared deep neural networks models: normalised complete FPs.} \label{fig:nets_complete}
 \end{figure*}
% ------------------------------------------------------------------------------------------------------------------------------

In the following experiment we evaluated the performance of our models (U--net, V--net and Res V--net) and the recent methods reported in Refs. \cite{ffdnet:zhang18, fpd:lin19, ffd:hao19}. We evaluated all the methods in the task of normalising FPs corrupted with Gaussian noise, speckle noise, illumination components variations and incomplete field of view. The training was conducted using $25,000$ patches for our models (U--net, V--net and ResV--net) and $40,000$ patches for the compared methods. The training set (patches) were randomly sampled over 30 of the total of 46 FPs. The remaining 16 FPs constitute the test set. For all the methods, we used the same training parameters, as described at the beginning of this section.

\begin{table}[t]
\centering
\caption{\bf Averaged errors over the training FPs (accuracy).}
\scalebox{0.92}{
\begin{tabular}{lccc}
\hline
	Model			&	MSE	$\times10^{-2}$	&	MAE	$\times10^{-1}$		&	 PSNR $\times10^{1}$ \\
	\hline
	x (input)		&	6.644				&	1.410				&	1.182 \\	
	FFD					&	5.008				&	1.938				&	1.302 \\	
	DCNN				&	4.048				&	1.047				&	1.416 \\	
	FPD					&	4.166				&	1.475				&	1.422 \\	
	U--net			&	0.846				&{\bf 0.587}	&	2.127 \\	
	V--net		 	&{\bf 0.764}	&	0.655				&{\bf 2.147} \\	
	Res V--net	& 0.824				&	0.788				&	2.087 \\
\hline
\end{tabular} }
	\label{tab:train_nets}
\end{table}

\begin{table}[t]
\centering
\caption{\bf Averaged errors over the test FPs (accuracy).}
\scalebox{0.92}{
\begin{tabular}{lccc}
\hline
	Model	&	MSE	$\times10^{-2}$	&	MAE	$\times10^{-1}$		&	 PSNR $\times10^{1}$ \\	
	\hline
	x (input)		&	6.501				&	1.393				&	1.194	\\
	FFD					&	5.251				&	2.012				&	1.283	\\
	DCNN				&	3.797				&	1.004				&	1.434	\\
	FPD					&	3.983				&	1.506				&	1.426	\\
	U--net			&	0.807				&{\bf 0.585}	&{\bf 2.138} \\
	V--net			&{\bf 0.765}	&	0.681				&{\bf	2.138} \\
	ResV-net		& 0.812				& 0.787				& 2.090	\\
	\hline
\end{tabular} }
	\label{tab:test_nets}
\end{table}

\begin{table}[h!]
\centering
\caption{\bf Averaged variations of the errors over the test FPs (precision).}
\scalebox{0.92}{
\begin{tabular}{lccc}
\hline
	Model		&	MSE		$\times10^{-5}$	&	MAE	$\times10^{-4}$		& $CV^2$ $\times10^{-3}$ \\	
	\hline
	x (input)		& 12.021			&	1.637				&	1.849 \\
	FFD					&	3.714				&	3.181				&	0.978 \\
	DCNN				&	7.959				&	1.861				&	1.998 \\
	FPD					& 18.296			&	3.623				&	3.484 \\
	U--net			&	1.737				&	0.664				&	2.151 \\
	V--net			&	0.702				&	0.476				& 0.918 \\
	Res V--net	&{\bf	0.313}	&{\bf 0.348}	&{\bf	0.385} \\
	\hline
\end{tabular} }
	\label{tab:test_nets_var}
\end{table}

Tables \ref{tab:train_nets}--\ref{tab:test_nets_var} summarise the experimental results. Table \ref{tab:train_nets} shows the averaged Mean Square Error (MSE), averaged Mean Absolute Error (MAE) and averaged Peak Signal to Noise Ratio (PSNR) over the full reconstructions of the 30 FPs used to generate the training set. Since we used patches for training, the full FPs were never seen for the networks. Table \ref{tab:test_nets} shows the averaged errors for the reconstructed 16 FPs used to generate the test set. Also, Table \ref{tab:test_nets_var} shows the averaged variance of the computed MSE and MAE of the errors in Table \ref{tab:test_nets}. The third column shows the square of the Coefficient of Variation ($CV^2$), where the relative standard deviation is defined as $CV = \sigma / | \mu |$. The $CV^2$ is a measure of the precision and repeatability of the results.

Finally, Figure \ref{fig:nets_patches} shows examples of reconstructed patches. Figure \ref{fig:nets_complete} shows examples of reconstructed full FPs. The procedure for reconstructing complete FPs was described in subsection 3.3. From a visual inspection of Figures \ref{fig:nets_patches} and \ref{fig:nets_complete}, one can note that the proposed networks produce the better results.

%------------------------------------------------------------------------------------------------------------------------
\subsection*{4.4  High-level and extremal noise}
\label{sec:noise}
%------------------------------------------------------------------------------------------------------------------------

In the following experiment we evaluated again the performance of the U--net, V--net, ResV--net, Fast V--net models versus and the methods reported in Refs. \cite{ffdnet:zhang18, fpd:lin19, ffd:hao19}. We evaluated all the methods in the task of normalising FPs corrupted with high levels of ESPI and Gaussian noise.

We use the second dataset of 180 images of size $320 \times 320$ for comparison purposes. In all models, the trainig set was the same and we maintain all hyper-parametes commmon to all models: ADAM optimiser, step size or learning rate of $\alpha$ in the range of $[1 \times 10^{-4}, 5 \times 10^{-4}]$, decay rate $= 10^{-3}$, batch size $= 32$. All the models were trained (ours and the reported DCNN, FFD and FPD) from zero using $40,000$ training patches of size $32 \times 32$, taken from the first 150 images. The remaining 30 images were used for testing.

All times correspond to the same computational resources. For the GFB method, we used a Python convolutional CPU implementation on an i5 Intel 3.7GHz and executing the process in one core. For all the neural network models, the training process is done with a Python--Tensorflow--Keras implementation, on a GPU NVIDIA GeForce 1080 Ti, and the process is executed used the NVIDIA 440 drivers and a CUDA 10.2 installation. The testing process uses a mixture between the i5 Intel (collecting and output of all covering patches) and the GPU (model inference). Table \ref{tab:times} summarises the computational time required for all models, both in the training stage and the inference process.

Tables \ref{tab:high_error}--\ref{tab:xtrm_error} summarise the experimental results. Table \ref{tab:high_error} shows the averaged Mean Square Error (MSE) and averaged Mean Absolute Error (MAE) over the full reconstructions of the 150 training and 30 testing FPs, for the ESPI and high-level Gaussian noise scenario. Similarly, Table \ref{tab:xtrm_error}  summarise the experimental results for the ESPI and extreme-level Gaussian noise scenario.

\begin{table}[t]
\centering
\caption{\bf Computation cost (time) of the evaluated models .}
\scalebox{0.9}{
\begin{tabular}{lccc}
\hline
	Model       & Training     & Inference   & Inference     \\
              &              & (per image) & efficiency    \\ \hline
	GFB		      & -----        & 387 s       & 0.155 img/min \\
	FFD         & 40-50 min    & 2.2 s       & 27 img/min    \\
	DCNN        & 30-35 min    & 2.9 s       & 20 img/min    \\
	FPD         & 140-150 min  & 10.1 s      & 6 img/min     \\
	U--net      & 13-15 min    & 2.5 s       & 24 img/min    \\
	V--net      & 40-50 min    & 6.1 s       & 10 img/min    \\
	ResV--net   & 50-60 min    & 6.0 s       & 10 img/min    \\
	Fast V--net & -----        & 0.024 s     & 2500 img/min  \\ \hline
\end{tabular} }
	\label{tab:times}
\end{table}

\begin{table}[t]
\centering
\caption{\bf Averaged errors over the training and testing FPs, high-noise level.}
%\scalebox{0.92}{
\begin{tabular}{lcccc}
\hline
	Model       &	MSE					&	MSE 		 		& MAE 			 	& MAE \\ 
	            &	train				&	test		 		& train 			& test \\ \hline
	FFD					&	1723.73			&	1747.28			&	26.7057			& 26.9246 \\	
	DCNN				&	252.45			&	253.71			&	9.1418 			& 9.1461 \\	
	FPD					&	129.10			&	145.03			&	5.2429			& 5.4363 \\	
	U--net			&	156.37			& 159.84			&	7.4497 			& 7.3633 \\	
	V--net		 	&{\bf 79.91}	&{\bf 97.88}	&{\bf 4.6833} &{\bf 4.8458} \\
	Res V--net	& 86.54				& 104.92			& 4.8618			& 5.1471 \\
	Fast V--net	& 104.44			&	110.95			&	5.7322			& 5.7431 \\
\hline
\end{tabular} 
	\label{tab:high_error}
\end{table}

\begin{table}[h!]
\centering
\caption{\bf Averaged errors over the training and testing FPs, extreme-noise level.}
%\scalebox{0.92}{
\begin{tabular}{lcccc}
\hline
	Model       &	MSE					&	MSE 		 		& MAE 			 	& MAE \\ 
	            &	train				&	test		 		& train 			& test \\ \hline
	FFD					&	2215.80			&	2248.44			&	29.8887			& 30.0743 \\	
	DCNN				&	1009.00			&	1038.59			&	18.6041			& 18.6464 \\	
	FPD					&	7839.25			&	841.23			&	13.1928			& 13.6363 \\	
	U--net			&	396.58			& 530.21			&	9.5679 			& 10.4250 \\	
	V--net		 	& 319.36    	& 474.76			& 8.8652			& 10.0145 \\	
	Res V--net	& 267.35			&	461.09			&{\bf 8.1307}	& 9.8830 \\
	Fast V--net	&{\bf 248.04}	&{\bf 390.51}	&	8.2544			&{\bf 9.6987} \\
\hline
\end{tabular} 
	\label{tab:xtrm_error}
\end{table}

\begin{figure*}[hp]
 \centering \centering \includegraphics[width=\linewidth]{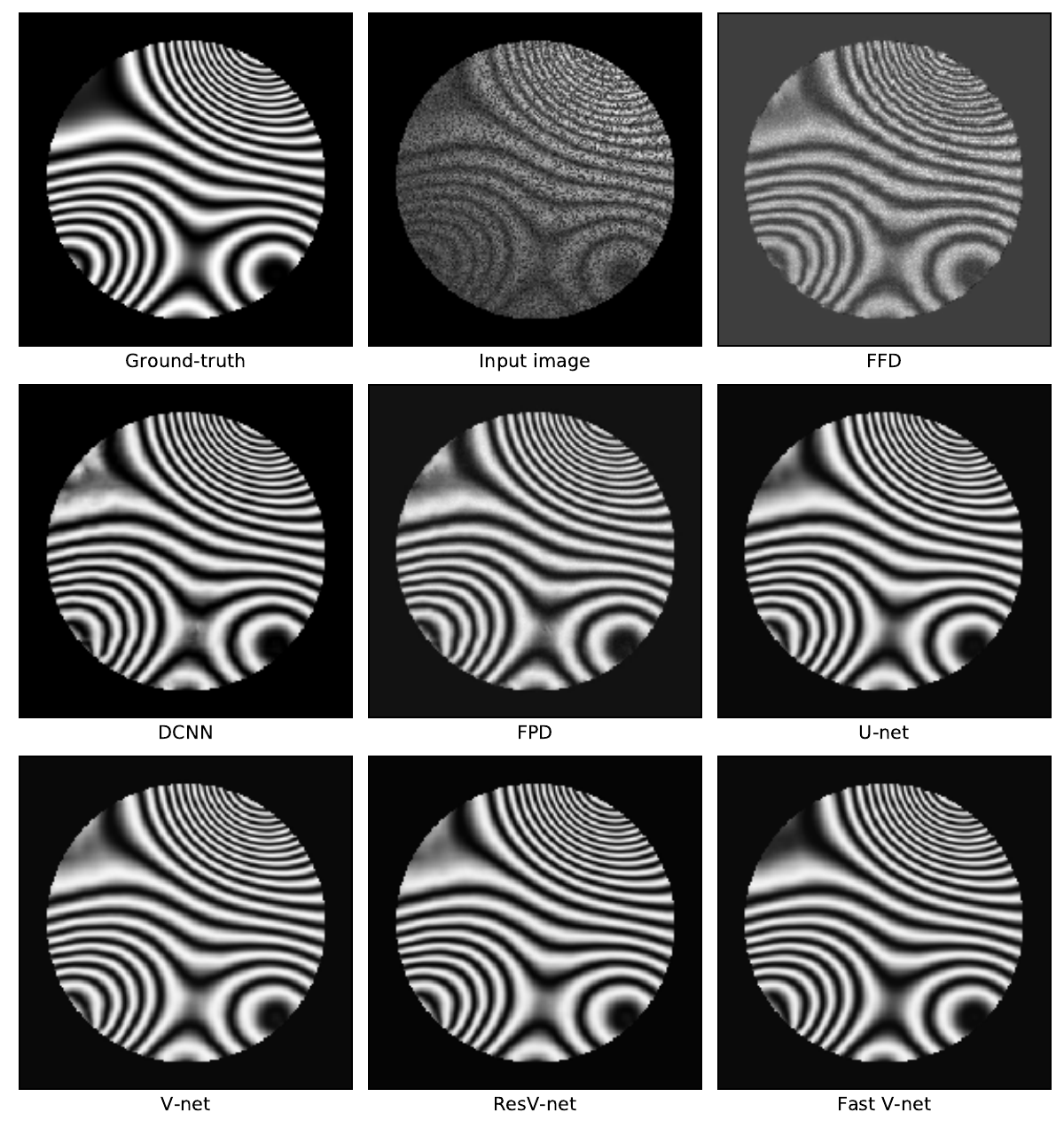}
 \caption{Results of the compared deep neural networks models: normalised FPs, in the scenario of ESPI and high-level speckle and Gaussian noise (test image \texttt{196}). Observe the better reconstruction obtained with all the V--net models.}
	 \label{fig:high_complete}
 \end{figure*}

\begin{figure*}[hp]
 \centering \centering \includegraphics[width=\linewidth]{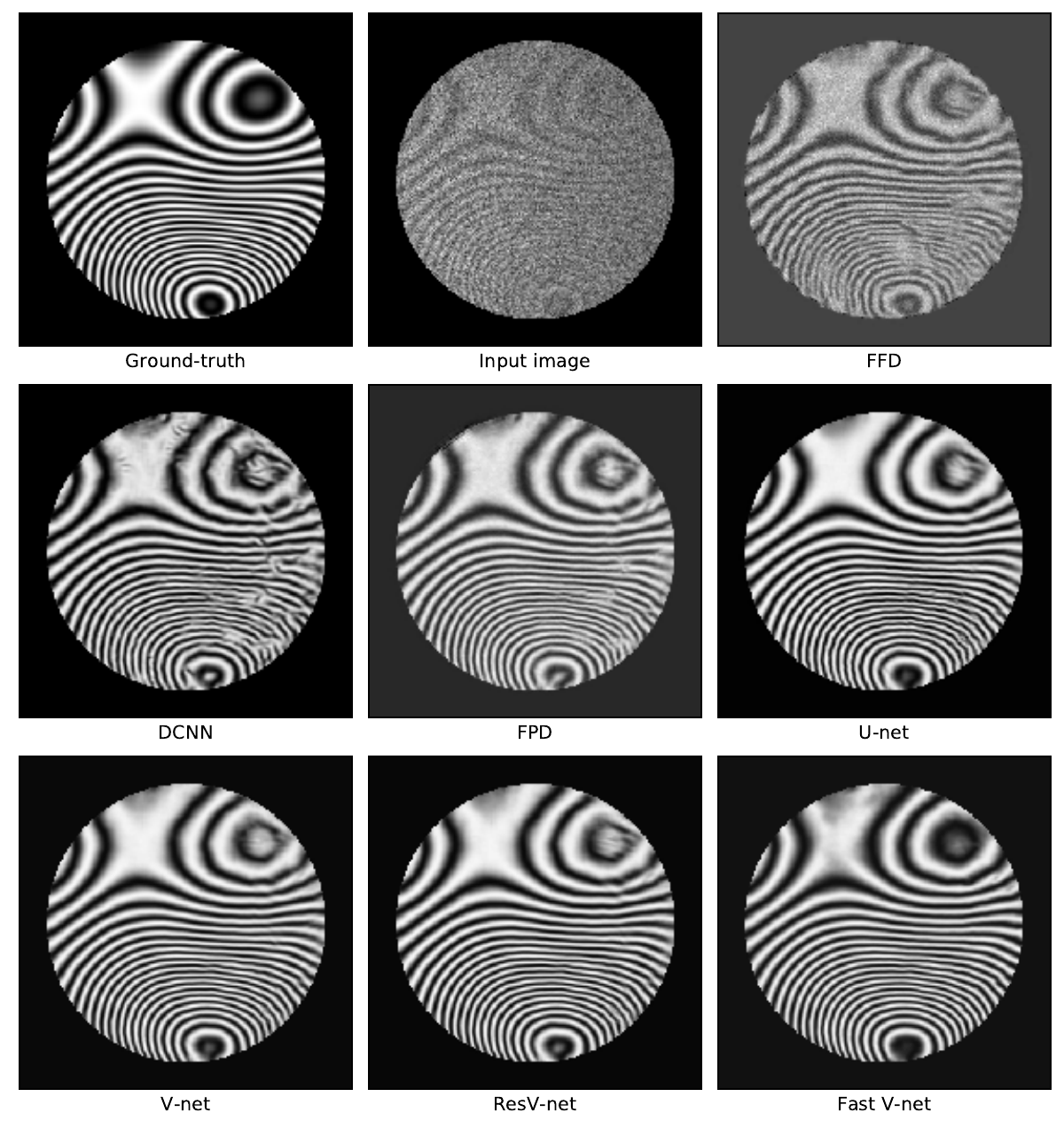}
 %\centering \centering \includegraphics[width=\linewidth]{xtrm/204.pdf}
 \caption{Results of the compared deep neural networks models: normalised FPs, in the scenario of ESPI and extreme-level spekle and Gaussian noise (test image \texttt{195}). Observe the better reconstruction obtained with all the V--net models.}
	 \label{fig:xtrm_complete}
 \end{figure*}

Figure \ref{fig:high_complete} shows examples of reconstructed full FPs, in the scenario in where the FPs with ESPI and high-level Gaussian noise. Similarly, Figure \ref{fig:xtrm_complete} shows examples of reconstructed full FPs, in the scenario in where the FPs with ESPI and extreme-level {\nobreak Gaussian} noise. As one can observe, again the proposed models generate better normalised reconstructions. Please refer to supplementary material for a better detail on the level of noise included in the synthetic data. Figs. \textbf{S1}-\textbf{S6} correspond to the high-level noise, Figs. \textbf{S7}-\textbf{S12} correspond to the extreme-level scenario. Figs. \textbf{S13}-\textbf{S18} show a comparison between all methods in the extreme-level case.

% ------------------------------------------------------------------------------------------------------------------------------

% ------------------------------------------------------------------------------------------------------------------------------

% ------------------------------------------------------------------------------------------------------------------------------

% ------------------------------------------------------------------------------------------------------------------------------

%------------------------------------------------------------------------------------------------------------------------
\subsection*{4.5  Evaluation on Real FPs}
\label{ssec:real}
%------------------------------------------------------------------------------------------------------------------------

\begin{figure}[ht]
 \centering \centering \includegraphics[width=0.9\linewidth]{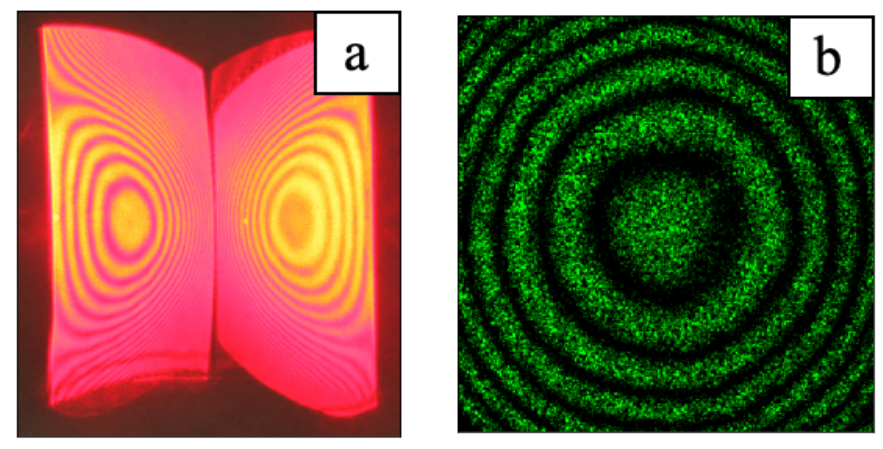}
 \caption{Real ESPI interferograms of a one--shot setup with a phase step equal $\pi$. (a) defocused and (b) focused.}
	 \label{fig:fp_reala}
 \end{figure}

\begin{figure}[h!]
 \centering \centering \includegraphics[width=\linewidth]{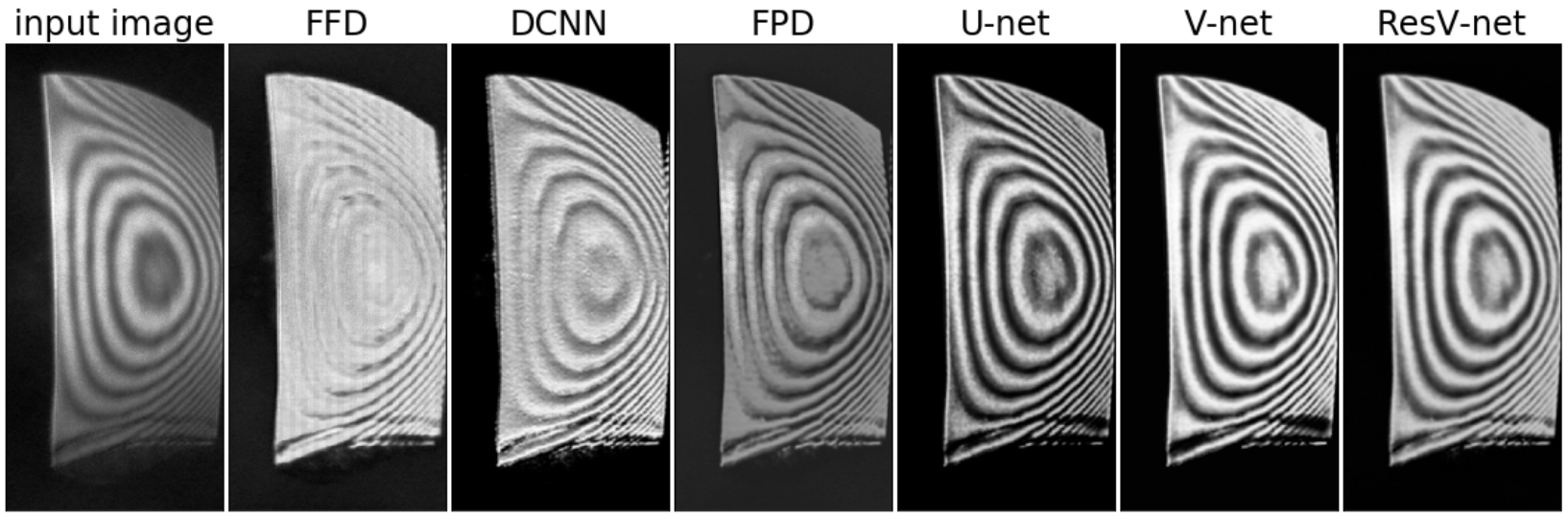}
 \caption{Results of the compared models with real interferometric FP in Figure \ref{fig:fp_reala}(a).}
	 \label{fig:fp_realb}
 \end{figure}

\begin{figure}[h!]
 \centering \centering \includegraphics[width=\linewidth]{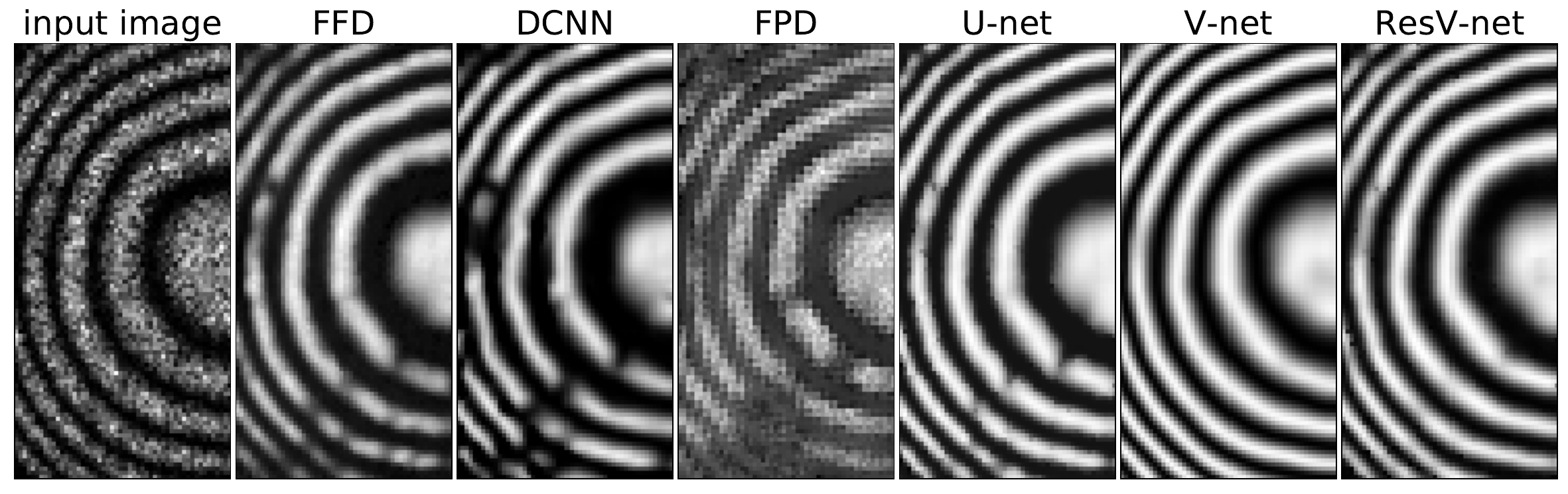}
 \caption{Results of the compared models with real interferometric FP in Figure \ref{fig:fp_reala}(b).}
	 \label{fig:fp_realc}
 \end{figure}

U--net, V--net, ResV--net and Fast V--net can be used to process real FPs, even when they were trainted with synthetic data. As an example, Figure \ref{fig:fp_reala} depicts a one--shot real interferometric ESPI-FP and  Figures \ref{fig:fp_realb}-\ref{fig:fp_realc} show the results of normalising the left section of the real interferogram by using the evaluated methods. We use the same models that were trained with simulated data in this experiment. The performance of all the evaluated methods is consistent with the obtained results when processing synthetic FPs.

% ------------------------------------------------------------------------------------------------------------------------------

% ------------------------------------------------------------------------------------------------------------------------------
% ------------------------------------------------------------------------------------------------------------------------------

% ------------------------------------------------------------------------------------------------------------------------------
% ------------------------------------------------------------------------------------------------------------------------------

%------------------------------------------------------------------------------------------------------------------------
\section*{6.  Discussion and conclusion}
\label{sec:conclu}
%------------------------------------------------------------------------------------------------------------------------

\subsection*{A.  Method limitations}

\begin{figure}[ht]
 \centering \centering \includegraphics[width=\linewidth]{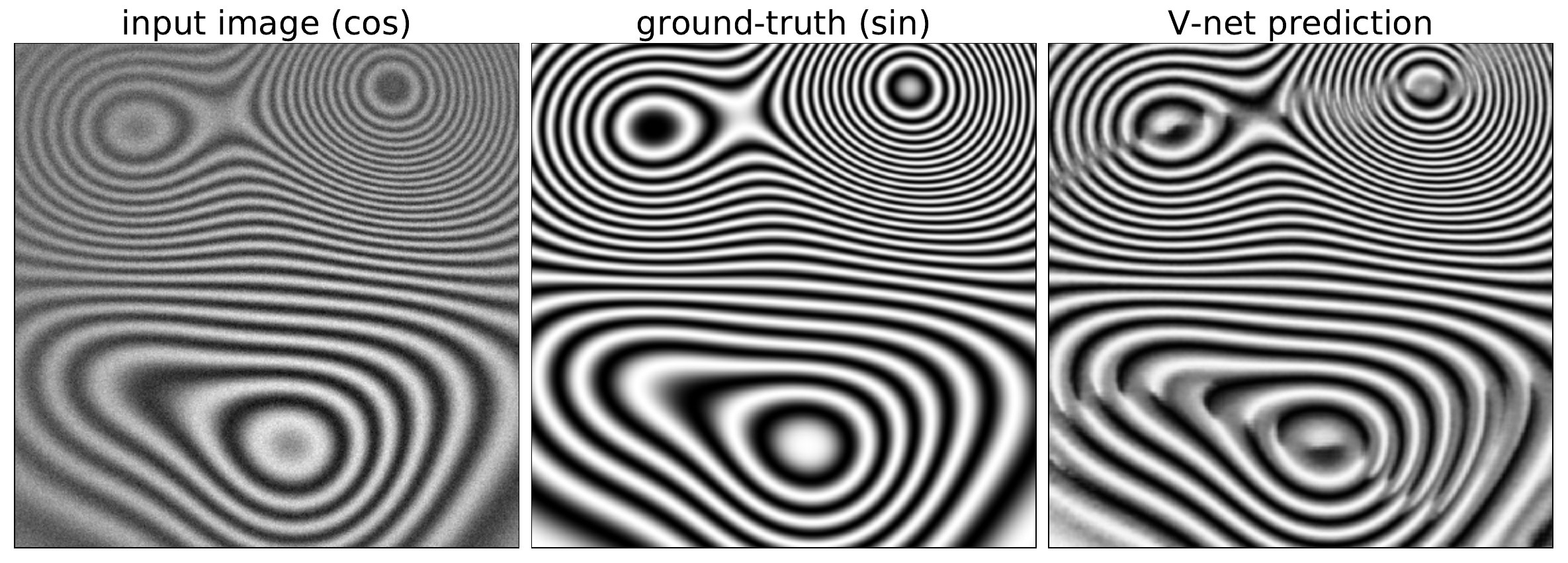}
 \caption{Computation of the normalised signal in quadrature. Note the global sign problem.}
  \label{fig:fp_complete_sin}
 \end{figure}

\begin{figure}[h!]
 \centering \centering \includegraphics[width=0.8\linewidth]{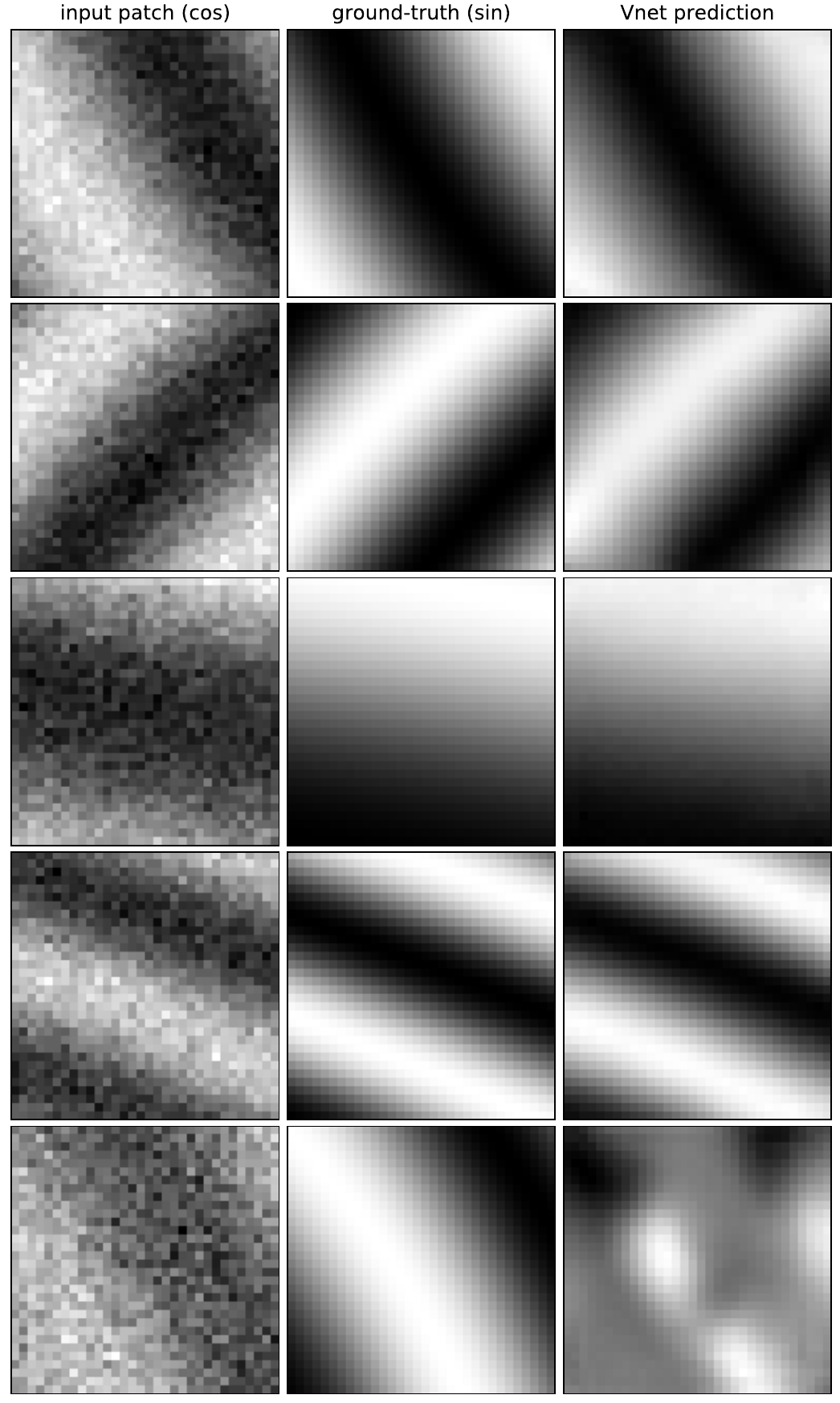}
 \caption{Computation of the normalised signal in quadrature at patch level. Note the sign problem in the patch at last row.}
	 \label{fig:fp_patches_sin}
 \end{figure}

U--net, V--net and ResV--net models base the FPs reconstruction by processing image patches with local information. Despite this is an advantage in terms of computational efficiency, there are some limitations for solving FP analysis problems associated with global information. To illustrate this point, we trained a V--net to estimate the quadrature normalised FP. Mathematically, we trained a V--net to estimate an operator $\Hop$, that given observations modelled by \eqref{eq:fp}, produces normalised FPs according to
 \begin{equation}
 	\label{eq:fp_sin}
	\hat x(p) = 1 + \sin\left( \phi(p) \right).
\end{equation} 
Figure \ref{fig:fp_complete_sin} shows a reconstructed quadrature FP where the ``global sign'' problem is evident. The problem actually occurs at patch level. There are patches for which the network can not infer the correct sign of the sine function, see reconstructed patch in last row in Figure \ref{fig:fp_patches_sin}. However, this sign change does not appear in arbitrary orientations, it seems that there exists a principal axis in the orientation domain where the sign changes systematically appear. 

We also observed that U--net, V--net and Res V--net models have limitations for filtering--out noise at regions with very low frequencies and low contrast; \ie at regions with low Signal to Noise Ratio (SNR). Those are regions where a visual inspection does not suggest a clear local dominant frequency. Moreover, we observed that ResV--net cannot completely remove the noise of figures in experiment of subsection 4.3. An explanation for this behaviour it that the ResV--net model assumes additive noise while our experimental data contains correlated noise (speckle). \\

On the other hand, the FastV--net model produces better results in extremal noise conditions, as can be noted in Table \ref{tab:xtrm_error} and Figure \ref{fig:xtrm_complete}. It can normalise FPs with good results, also in regions of high-frequency, even in the boundary of Nyquist phenomenon. Again, bad contrast and bad frequency regions bound the quality of the reconstruction.

% ------------------------------------------------------------------------------------------------------------------------------

% ------------------------------------------------------------------------------------------------------------------------------

% ------------------------------------------------------------------------------------------------------------------------------
% ------------------------------------------------------------------------------------------------------------------------------

\subsection*{B.  Conclusions}

The proposed normalisation--denoising method for FPs
 is based on the deep learning paradigm. Under this paradigm, one trains a neural network (universal approximator) that estimates an appropriated transformation between observations (corrupted inputs) and outputs. In particular, our solution builds upon a deep auto--encoder that produces results with very small errors. Our results show that the proposed U--net and V--net schemes can be applied to real FP images, in which the image's corruption irregularities correspond to high noise levels, illumination component variations or an incomplete field of view; see Figures \ref{fig:gfb} and \ref{fig:nets_complete}. Our models can process real FPs, even if we train the networks with simulated data.

We observed that DNNs can compute reconstructions with lower error than GFBs near the boundary of the field of view; see Figure \ref{fig:gfb}. We found that V--net produces higher quality reconstruction for FPs for higher levels of noise and pupils than U--net and the compared methods. In our opinion, the reason is that the V--net filter distribution across the layers is designed to retain more details, as opposed to U--net which is designed to segment images.

We believe that the evaluated methods are a new research branch for developing more sophisticated FPs analysis methods, that based on deep neural networks, can compute solutions in a front--to--end strategy. It could be interesting to design and implement specific deep networks architectures for solving challenging problems in FP analysis; \eg, the phase recovery from a single interferogram with closed fringes and the quadrature FP computation.

%------------------------------------------------------------------------------------------------------------------------
%\section*{Acknowledgments} 
%------------------------------------------------------------------------------------------------------------------------

%\noindent The authors thank Adonai Gonz\'alez and V\'{i}ctor Flores for providing the data for Figure \ref{fig:fp_reala}. This research was supported in part by Conacyt, Mexico (A1-S-43858 research grant and A.R.F. PhD. studies scholarship) and the NVIDIA Academic program.

%------------------------------------------------------------------------------------------------------------------------
\section*{Disclosures} 
%------------------------------------------------------------------------------------------------------------------------

\textbf{Funding.} Concejo Nacional de Ciencia y Tecnología (CONACYT) Mexico (Grant A1-S-43858 , PhD scholarship). \\
\textbf{Disclosures.} MR: Nvidia Academic Program (F). \\

\noindent See Supplement 1 for supporting content.

% ------------------------------------------------------------------------------------------------------------------------------

%\section*{References}
% ------------------------------------------------------------------------------------------------------------------------------

\bibliographystyle{osajnl}
%\biboptions{sort&compress}

%------------------------------------------------

%\bibliography{SLEF}

\begin{thebibliography}{1}

%\expandafter\ifx\csname url\endcsname\relax
	%\def\url#1{\texttt{#1}}\fi
%\expandafter\ifx\csname urlprefix\endcsname\relax\def\urlprefix{URL }\fi
%\expandafter\ifx\csname href\endcsname\relax
	%\def\href#1#2{#2} \def\path#1{#1}\fi

\bibitem{quiroga:03}
	J.A.~Quiroga, M.~Serv\'{i}n,
	\href{https://www.sciencedirect.com/science/article/abs/pii/S0030401803018017}{Isotropic $n$-dimensional fringe pattern normalization}, Optics Communications 224 (2003) 221--227.
\newblock \href {https://doi.org/10.1016/j.optcom.2003.07.014}
	{\path{doi.org/10.1016/j.optcom.2003.07.014}}.

\bibitem{quiroga:01}
	J.A.~Quiroga, J.A.~G\'{o}mez-Pedrero, A.~Garc\'{i}a-Botella,
	\href{https://www.sciencedirect.com/science/article/abs/pii/S0030401801014407}{Algorithm for fringe pattern normalization}, Optics Communications 197 (2001) 43--51.
\newblock \href {https://doi.org/10.1016/S0030-4018(01)01440-7}
	{\path{doi.org/10.1016/S0030-4018(01)01440-7}}.

\bibitem{marroquin:97}
	J.~L.~Marroqu\'{i}n, M.~Serv\'{i}n, and R.~Rodriguez-Vera, {Adaptive quadrature filters and the recovery of phase from fringe pattern images,} J. Opt. Soc. Am. A 14 (1997) 1742--1753.

\bibitem{servin:98}
	M.~Serv\'{i}n, J.L.~Marroqu\'{i}n, D.~Malacara, F.~Cuevas,
	\href{https://www.osapublishing.org/ao/abstract.cfm?uri=ao-37-10-1917}{Phase unwrapping with a regularized phase-tracking system }, Appl. Optics 37(10) (1998) 1917--1923.
\newblock \href {https://doi.org/10.1364/AO.37.001917}
	{\path{doi.org/10.1364/AO.37.001917}}.
		
\bibitem{servin:01}
	M.~Serv\'{i}n, J.L.~Marroqu\'{i}n, and F.~J.~Cuevas, {Fringe-follower regularized phase tracker for demodulation of closed-fringe interferograms,} J. Opt. Soc. Am. A 18 (2001) 689--695.

\bibitem{rivera:05}
	M.~Rivera,{Robust phase demodulation of interferograms with open or closed fringes,} J. Opt. Soc. Am. A 22 (2005) 1170--1175.

\bibitem{Flores:2020}
V.~H.~Flores, A.~Reyes-Figueroa, C.~Carrillo-Delgado, M.~Rivera,
	\href{https://www.sciencedirect.com/science/article/abs/pii/S0030399219317128}{Two-step phase shifting algorithms: Where are we?}, Opt. and Laser Technology 126 (2020) 1--13.
\newblock \href {https://doi.org/10.1016/j.optlastec.2020.106105}
  {\path{doi.org/10.1016/j.optlastec.2020.106105}}.
 
\bibitem{gorthi:fp10}
S.~S. Gorthi, P.~Rastogi,
  \href{http://www.sciencedirect.com/science/article/pii/S0143816609002164}{Fringe
  projection techniques: Whither we are?}, Optics and Lasers in Engineering
  48~(2) (2010) 133--140, Fringe Projection Techniques.
\newblock \href{https://doi.org/10.1016/j.optlaseng.2009.09.001}
  {\path{doi.org/10.1016/j.optlaseng.2009.09.001}}.

\bibitem{juarez:fprev15}
R.~Juarez-Salazar, F.~Guerrero-Sanchez, C.~Robledo-Sanchez,
  \href{http://ao.osa.org/abstract.cfm?URI=ao-54-17-5364}{Theory and algorithms
  of an efficient fringe analysis technology for automatic measurement
  applications}, Appl. Opt. 54~(17) (2015) 5364--5374.
\newblock \href {http://dx.doi.org/10.1364/AO.54.005364}
  {\path{doi.org/10.1364/AO.54.005364}}.

\bibitem{kemao2007two}
Q.~Kemao,
	\href{https://www.sciencedirect.com/science/article/pii/S0143816606000455}{Two-dimensional
	windowed {F}ourier transform for fringe pattern analysis: principles, applications and
	implementations}, Optics and Lasers	in Engineering 45~(2) (2007) 304--317.
\newblock \href {https://doi.org/10.1016/j.optlaseng.2005.10.012}
  {\path{doi.org/10.1016/j.optlaseng.2005.10.012}}.

\bibitem{huang:waveletFP10}
L.~Huang, Q.~Kemao, B.~Pan, A.~K. Asundi,
	\href{https://www.sciencedirect.com/science/article/pii/S0143816609000840}{Comparison of {F}ourier transform, windowed {F}ourier transform, and wavelet transform methods for phase extraction from a single fringe pattern in fringe projection profilometry}, Optics and Lasers in Engineering 48~(2) (2010) 141--148, fringe Projection Techniques.
\newblock \href {https://doi.org/10.1016/j.optlaseng.2009.04.003}
  {\path{doi.org/10.1016/j.optlaseng.2009.04.003}}.

\bibitem{zhang:wft12}
Z.~Zhang, Z.~Jing, Z.~Wang, D.~Kuang,
  \href{http://www.sciencedirect.com/science/article/pii/S0143816612000760}{Comparison
  of {F}ourier transform, windowed {F}ourier transform, and wavelet transform
  methods for phase calculation at discontinuities in fringe projection
  profilometry}, Optics and Lasers in Engineering 50~(8) (2012) 1152--1160.
\newblock \href {https://doi.org/10.1016/j.optlaseng.2012.03.004}
  {\path{doi.org/10.1016/j.optlaseng.2012.03.004}}.

\bibitem{rivera:twostep16}
M.~Rivera, O.~Dalmau, A.~Gonzalez, F.~Hernandez-Lopez,
  \href{http://www.sciencedirect.com/science/article/pii/S0143816616300471}{Two-step fringe pattern analysis with a {G}abor filter bank}, Opt. Lasers Eng. 85
  (2016) 29--37.
\newblock \href {https://doi.org/10.1016/j.optlaseng.2016.04.014}
  {\path{doi.org/10.1016/j.optlaseng.2016.04.014}}.

\bibitem{dalmau:twostep16}
O.~Dalmau, M.~Rivera, A.~Gonzalez,
  \href{http://www.sciencedirect.com/science/article/pii/S0030401816302279}{Phase shift estimation in interferograms with unknown phase step}, Opt. Commun. 372
  (2016) 37--43.
\newblock \href {https://doi.org/10.1016/j.optcom.2016.03.063}
  {\path{doi.org/10.1016/j.optcom.2016.03.063}}.

\bibitem{rivera:transient18}
M.~Rivera,
  \href{http://www.sciencedirect.com/science/article/pii/S0143816617311910}{Robust
  fringe pattern analysis method for transient phenomena}, Optics and Lasers in
  Engineering 108 (2018) 19--27.
\newblock \href {https://doi.org/10.1016/j.optlaseng.2018.03.013}
  {\path{doi.org/10.1016/j.optlaseng.2018.03.013}}.

\bibitem{kurkova92:approxNN}
V.~K{\r u}rkov{\'a},
  \href{http://www.sciencedirect.com/science/article/pii/0893608092900128}{Kolmogorov's theorem and multilayer neural networks}, Neural Networks 5~(3) (1992) 501--506.
\newblock \href {https://doi.org/10.1016/0893-6080(92)90012-8}
  {\path{doi.org/10.1016/0893-6080(92)90012-8}}.

\bibitem{cuevas00:nnfringes}
F.~Cuevas, M.~Servin, O.~Stavroudis, R.~Rodriguez-Vera,
	\href{https://www.sciencedirect.com/science/article/pii/S0030401800007653}{Multi-layer
	neural network applied to phase and depth recovery from fringe patterns}, Optics
  Communications 181~(4) (2000) 239--259.
\newblock \href {https://doi.org/10.1016/S0030-4018(00)00765-3}
  {\path{doi.org/10.1016/S0030-4018(00)00765-3}}.

%\bibitem{goodfellow2016deep},I.~Goodfellow, Y.~Bengio, A.~Courville, Deep learning, MIT Press (2016)

\bibitem{Hinton504}
G.~E. Hinton, R.~R. Salakhutdinov,
  \href{http://science.sciencemag.org/content/313/5786/504}{Reducing the
  dimensionality of data with neural networks}, Science 313~(5786) (2006)
  504--507.
\newblock \href {http://dx.doi.org/10.1126/science.1127647}
  {\path{doi.org/10.1126/science.1127647}}.

\bibitem{nair10:relu}
V.~Nair, G.~E.~Hinton. Rectified linear units improve restricted Boltzmann machines. In Proc of 27th Int. Conf. on Machine Learning (ICML) 2010, pp. 807--814.

\bibitem{ronnenberg:Unet15}
O.~Ronneberger, P.~Fischer, T.~Brox,
	\href{https://link.springer.com/chapter/10.1007/978-3-319-24574-4_28}{U-net:
	Convolutional networks for biomedical image segmentation}, in:
	N.~Navab, J.~Hornegger, W.~M. Wells, A.~F. Frangi (Eds.), Medical Image
	Computing and Computer-Assisted Intervention -- MICCAI 2015, Springer
	International Publishing, Cham, 2015, pp. 234--241.
\newblock \href {http://doi.org/10.1007/978-3-319-24574-4_28}
  {\path{doi.org/10.1007/978-3-319-24574-4_28}}.
	
\bibitem{fullyconvnet}
J.~Long, E.~Shelhamer, T.~Darrell,
	\href{https://ieeexplore.ieee.org/document/7298965}{Fully convolutional
	networks for semantic segmentation}, in: 2015 IEEE Conference on Computer
	Vision and Pattern Recognition (CVPR), 2015, pp. 3431--3440.
\newblock \href {http://dx.doi.org/10.1109/CVPR.2015.7298965}
  {\path{doi.org/10.1109/CVPR.2015.7298965}}.

%\bibitem{jin:dnn_inverse17}
%K.~H. Jin, M.~T. McCann, E.~Froustey, M.~Unser,
%	\href{https://ieeexplore.ieee.org/document/7949028}{Deep convolutional neural
%	network for inverse problems in imaging}, IEEE Transactions on Image
%	Processing 26~(9) (2017) 4509--4522.
%\newblock \href {http://dx.doi.org/10.1109/TIP.2017.2713099}
%	{\path{doi.org/10.1109/TIP.2017.2713099}}.

\bibitem{he:resnet16}
K.~He, X.~Zhang, S.~Ren, J.~Sun,
	\href{https://ieeexplore.ieee.org/document/7780459}{Deep residual learning for image
	recognition}, in: 2016 IEEE Conference on Computer Vision and Pattern Recognition (CVPR),
  2016, pp. 770--778.
\newblock \href {http://dx.doi.org/10.1109/CVPR.2016.90}
  {\path{doi.org/10.1109/CVPR.2016.90}}.

\bibitem{renteria:2020}
O.I.~Renteria-Vidales, J.C.~Cuevas-Tello, A.~Reyes-Figueroa, M.~Rivera,
	\href{https://link.springer.com/chapter/10.1007/978-3-030-49076-8_21}{ModuleNet: A Convolutional Neural Network for Stereo Vision}, in Pattern Recognition. MCPR 2020. Lecture Notes in Comp. Sc. 12088, Springer, (2020) 219--228,
\newblock \href {https://doi.org/10.1007/978-3-030-49076-8_21}
  {\path{doi.org/10.1007/978-3-030-49076-8_21}}.

%\bibitem{mao:dnnrestor}
%X.~Mao, C.~Shen, Y.-B. Yang,
%  \href{http://papers.nips.cc/paper/6172-image-restoration-using-very-deep-convolutional-encoder-decoder-networks-with-symmetric-skip-connections.pdf}{Image
%  restoration using very deep convolutional encoder-decoder networks with
%  symmetric skip connections}, in: D.~D. Lee, M.~Sugiyama, U.~V. Luxburg,
%  I.~Guyon, R.~Garnett (Eds.), Advances in Neural Information Processing
%  Systems 29, Curran Associates, Inc., 2016, pp. 2802--2810.

%\bibitem{srivastava14:dropout}
%N.~Srivastava, G.~Hinton, A.~Krizhevsky, I.~Sutskever, R.~Salakhutdinov,
%  \href{http://jmlr.org/papers/v15/srivastava14a.html}{Dropout: A simple way to
%  prevent neural networks from overfitting}, Journal of Machine Learning
%  Research 15 (2014) 1929--1958.

\bibitem{posix:rotenberg60}
A.~Rotenberg,
	\href{https://dl.acm.org/citation.cfm?doid=321008.321019}{A new pseudo-random
	number generator}, J. ACM 7 (1) (1960) 75--77.
\newblock \href {http://doi.acm.org/10.1145/321008.321019}
	{\path{doi.acm.org/10.1145/321008.321019}}.

\bibitem{rbf:broomhead88}
D.~Broomhead, D.~Lowe,
	\href{https://www.complex-systems.com/abstracts/v02_i03_a05/}{Multivariable functional interpolation and adaptive networks}, Complex Systems 2 (1988) 321--355.

\bibitem{adam:kingma15}
D.~P.~Kingma, J.~Ba,
	\href{https://arxiv.org/abs/1412.6980}{Adam: A method for stochastic optimization},
	in 3rd International Conference for Learning Representations, San Diego, 2015 (2014).
\newblock \href {http://arxiv.org/abs/1412.6980}
	{\path{arxiv.org/abs/1412.6980}}.

\bibitem{fpd:lin19}
B.~Lin, S.~Fu, C.~Zhang, F.~Wang, Y.~Li,
	\href{https://arxiv.org/abs/1901.00361}{Optical fringe patterns filtering based
	on multi-stage convolution neural network}, Preprint.
\newblock \href {https://arxiv.org/abs/1901.00361}
	{\path{arxiv.org/abs/1901.00361}}.

\bibitem{dcnn:yan19}
K.~Yan, Y.~Yu, C.~Huang, L.~Sui, K.~Qian, A.~Asundi,
	\href{https://www.sciencedirect.com/science/article/pii/S0030401818311076}{Fringe
	pattern denoising based on deep learning}, Optics Communications 437 (2019) 148--152.
\newblock \href {https://doi.org/10.1016/j.optcom.2018.12.058}
	{\path {doi.org/10.1016/j.optcom.2018.12.058}}.

\bibitem{ffd:hao19}
F.~Hao, C.~Tang, M.~Xu, Z.~Lei,
	\href{https://www.osapublishing.org/ao/abstract.cfm?uri=ao-58-13-3338}{Batch
	denoising of {ESPI} fringe patterns based on convolutional neural network},
	Appl. Opt. 58 (13) (2019) 3338--3346.
\newblock \href {https://doi.org/10.1364/AO.58.003338}
	{\path{doi.org/10.1364/AO.58.003338}}.

\bibitem{ffdnet:zhang18}
K.~Zhang, W.~Zuo, L.~Zhang,
	\href{https://ieeexplore.ieee.org/document/8365806?}{{FFDN}et: Toward a fast
	and flexible solution for {CNN}-based image denoising}, IEEE Transactions
	on Image Processing 27 (2018) 4608--4622.
\newblock \href {https://doi.org/10.1109/TIP.2018.2839891}
	{\path {doi.org/10.1109/TIP.2018.2839891}}.

\bibitem{Kemao:04}
	Q.~Kemao,
	\href{https://www.osapublishing.org/ao/abstract.cfm?uri=ao-43-13-2695}{Windowed
	{F}ourier transform for fringe pattern analysis}, Appl. Opt. 43 (13) (2004) 2695--2702.
\newblock \href {https://doi.org/10.1364/AO.43.002695}
	{\path{doi.org/10.1364/AO.43.002695}}.

\end{thebibliography}

\end{document}